\documentclass[aps, prd, twocolumn, amsmath, floats,floatfix, superscriptaddress, nofootinbib, showpacs]{revtex4}
\usepackage{amssymb}
\usepackage{amsmath}
\usepackage{verbatim}
\usepackage{mathrsfs}
\usepackage{amsfonts}
\usepackage{latexsym}
\usepackage{epsfig}
\usepackage{epstopdf}
\usepackage{color}
\usepackage{graphicx,subfigure}
\usepackage{units}
\usepackage{slashbox}

\newcommand{\Real}{\mathbb{R}}
\newcommand{\Complex}{\mathbb{C}}
\newcommand{\re}{\mbox{Re}}
\newcommand{\im}{\mbox{Im}}
\newcommand{\arcsinh}{\mbox{arcsinh}}

\newtheorem{proposition}{Proposition}
\newcommand{\proof}{\noindent {\bf Proof. }}
\newcommand{\qed}{\hfill $\fbox{\hspace{0.3mm}}$ \vspace{.3cm}} %End of proof.

\begin{document}

%%%%%%%%%%%%%%%%%%
%%%   MACROS   %%%
%%%%%%%%%%%%%%%%%%

\long\def\symbolfootnote[#1]#2{\begingroup%
\def\thefootnote{\fnsymbol{footnote}}\footnote[#1]{#2}\endgroup}

%%%%%%%%%%%%%%%%%%%%
%%%   TITLE  and AUTHORS   %%%
%%%%%%%%%%%%%%%%%%%%
\title{Schwarzschild scalar wigs: spectral analysis and late time behavior}

\author{Juan Barranco} 
%\email[]{barranco@astroscu.unam.mx}
\affiliation{Departamento de F\'isica, Divisi\'on de Ciencias e Ingenier\'ias, Campus Le\'on, 
Universidad de Guanajuato, Le\'on 37150, M\'exico}

\author{Argelia Bernal} 
%\email[]{argelia.bernal@nucleares.unam.mx}
\affiliation{\'Area Acad\'emica de Matem\'aticas y F\'isica, Universidad
Aut\'onoma del Estado de Hidalgo, Carretera Pachuca-Tulancingo Km. 4.5, C.P. 42184,
Pachuca, Hidalgo, M\'exico}

\author{Juan Carlos Degollado} 
%\email[]{jdaza@astroscu.unam.mx}
\affiliation{Departamento de F\'isica da Universidade de Aveiro and I3N,
Campus de Santiago, 3810-183 Aveiro, Portugal}

\author{Alberto Diez-Tejedor}
%\email[]{alberto.diez@nucleares.unam.mx}
\affiliation{Departamento de F\'isica, Divisi\'on de Ciencias e Ingenier\'ias, Campus Le\'on, 
Universidad de Guanajuato, Le\'on 37150, M\'exico}
\affiliation{SCIPP and Department of Physics, University of California,
Santa Cruz, CA, 95064, USA}

\author{Miguel Megevand}
%\email[]{megevand@nucleares.unam.mx}
\affiliation{Department of Physics and Astronomy, Louisiana State University,
  Baton Rouge, LA 70803-4001, USA}
\affiliation{Instituto de Ciencias Nucleares, Universidad Nacional
  Aut\'onoma de M\'exico, Circuito Exterior C.U., A.P. 70-543,
  M\'exico D.F. 04510, M\'exico}  

\author{Miguel Alcubierre}
%\email[]{malcubi@nucleares.unam.mx}
\affiliation{Instituto de Ciencias Nucleares, Universidad Nacional
  Aut\'onoma de M\'exico, Circuito Exterior C.U., A.P. 70-543,
  M\'exico D.F. 04510, M\'exico}

\author{Dar\'{\i}o N\'u\~nez}
%\email[]{nunez@nucleares.unam.mx}
\affiliation{Instituto de Ciencias Nucleares, Universidad Nacional
  Aut\'onoma de M\'exico, Circuito Exterior C.U., A.P. 70-543,
  M\'exico D.F. 04510, M\'exico}

\author{Olivier Sarbach}
%\email[]{sarbach@ifm.umich.mx}
\affiliation{Instituto de F\'{\i}sica y Matem\'aticas, Universidad
Michoacana de San Nicol\'as de Hidalgo, Edificio C-3, Ciudad
Universitaria, 58040 Morelia, Michoac\'an, M\'exico}

%%%%%%%%%%%%%%%%
%%%   DATE   %%%
%%%%%%%%%%%%%%%%

\date{\today}

%%%%%%%%%%%%%%%%%%%%
%%%   ABSTRACT   %%%
%%%%%%%%%%%%%%%%%%%%

\begin{abstract}  
Using the Green's function representation technique, the late time behavior of
localized scalar field distributions on Schwarzschild spacetimes is studied.
Assuming arbitrary initial data we perform a spectral analysis, computing the
amplitude of each excited quasi-bound mode without the necessity of performing
dynamical evolutions. The resulting superposition of modes is compared with
a traditional numerical evolution with excellent agreement; therefore, we have
an efficient way to determine final black hole {\it wigs}. The astrophysical
relevance of the quasi-bound modes is discussed in the context of scalar field
dark matter models and the axiverse.
\end{abstract}

%%%%%%%%%%%%%%%%
%%%   PACS   %%%
%%%%%%%%%%%%%%%%


%%%%%%%%%%%%%%%%%%%%%%
%%%   MAKE TITLE   %%%
%%%%%%%%%%%%%%%%%%%%%%

\maketitle

%%%%%%%%%%%%%%%%%%%%%%%%
%%%   INTRODUCTION   %%%
%%%%%%%%%%%%%%%%%%%%%%%%

%%%%%%%%%%%%%%%%%%%%%%%%%%%%%%%%%%%%%%%%%%%%%%
\section{Introduction}
\label{sec:introduction}
%%%%%%%%%%%%%%%%%%%%%%%%%%%%%%%%%%%%%%%%%%%%%%

Scalar fields show up in many areas of physics. In the standard model of
particle physics they are necessary to give particles their observed
mass. At the effective level, the charged pions mediating
the strong nuclear interaction are described by a scalar field. 
Another example is provided by the axion, introduced in order to
resolve the strong CP problem in quantum chromodynamics. Generically, a plethora
of scalar fields emerge in the {\it axiverse} from super-string and higher
dimensional theories~\cite{Arvanitaki:2009fg,Acharya:2010zx,Arvanitaki:2010sy,
Marsh:2011gr,Kodama:2011zc,Macedo:2013qea,Marsh:2013taa,Tashiro:2013yea}. 
In a cosmological context, a scalar field could guide an early phase of exponential
expansion, also represent the dark energy, 
and possibly even the dark matter in the present Universe~\cite{Turner:1983he,
Sin:1992bg, Peebles:1999fz, Peebles:2000yy, Sahni:1999qe, Hu:2000ke,
Matos:2000ng, Matos:2000ss, Arbey:2001qi, Matos:2003pe, Lee:2008jp,
Sikivie:2009qn, Marsh:2010wq, Lundgren:2010sp, Su:2010bj, Briscese:2011ka,
Harko:2011xw,
Lora:2011yc, GonzalezMorales:2012uw, Rindler-Daller:2013zxa}.

On the other hand, black holes are a natural consequence of general relativity.
They appeared for the first time as a purely mathematical result, but they were
later recognized as real astrophysical objects. Presently, we believe black
holes represent the final fate of sufficiently massive stars, and supermassive
black holes probably reside at the center of most
galaxies~\cite{Ferrarese:2000se}.

Since both scalar fields and black holes seem to inhabit our Universe, a natural question arises: How 
do scalar fields and black holes interact with each other? Static, spherically symmetric,
asymptotically flat black holes with a non-trivial scalar field distribution cannot exist in 
nature~\cite{Bekenstein:1995un, Pena:1997cy}; however, these results do not rule out extremely long lived
transient phenomena~\cite{Burt:2011pv, Barranco:2012qs}.

The massive Klein-Gordon equation on a black hole background has been studied
extensively. For a Schwarzschild black hole Ternov {\it et al.}~\cite{Ternov:1978gq}
found the existence of quasi-bound modes (also called quasi-stationary modes or
quasi-resonances), describing scalar field configurations surrounding the black
hole for some period of time; see also Refs.~\cite{Ohashi:2004wr,Grain:2007gn} for later work on these states.
Similar modes were found on a Kerr black hole
background~\cite{Ternov:1978gq, Detweiler:1980uk, Dolan:2007mj}. In the rotating case, there is a specific range 
of parameters for which these modes are exponentially growing in time, implying that a massive scalar 
field on a rotating Kerr black hole is unstable.
See Refs.~\cite{Beyer:2000fz, Beyer:2011py, Beyer:2012va, Shlapentokh-Rothman:2013ysa} for rigorous results on this 
instability, and~\cite{Witek:2012tr, Dolan:2012yt, Okawa:2014nda} for recent numerical studies. Based on Leaver's
continuous fraction method~\cite{Leaver:1985ax}, Konoplya and Zhidenko~\cite{Konoplya:2004wg}
computed the quasi-normal frequencies of a massive scalar field on a Schwarzschild spacetime 
background, and mentioned conditions under which 
quasi-bound states could arise. For a
generalization to the Kerr-Newman case, see Ref.~\cite{Furuhashi:2004jk}. 
Finally, we mention that Dolan and collaborators~\cite{Lasenby:2002mc,Rosa:2011my} generalized 
the computation of quasi-bound modes to different
types of fields, vectorial or fermionic.

Motivated by the possibility of describing the dark matter in the galactic halo
by the coherent excitation of an ultra-light scalar field~\cite{Turner:1983he,
Sin:1992bg, Peebles:1999fz, Peebles:2000yy, Sahni:1999qe, Hu:2000ke,
Matos:2000ng, Matos:2000ss, Arbey:2001qi, Matos:2003pe, Lee:2008jp,
Sikivie:2009qn, Marsh:2010wq, Lundgren:2010sp, Su:2010bj, Briscese:2011ka,
Harko:2011xw,
Lora:2011yc, GonzalezMorales:2012uw, Rindler-Daller:2013zxa}, including the supermasive black 
hole~\cite{UrenaLopez:2002du, CruzOsorio:2010qs, UrenaLopez:2011fd, Guzman:2012jc}, 
in previous work we analyzed the time scale for the quasi-bound states on a
supermassive, nonrotating black hole background~\cite{Burt:2011pv, Barranco:2012qs};
see ~\cite{Zhou:2013dra} for a recent generalization to the Dirac field.
In particular, we showed in Ref.~\cite{Barranco:2012qs} that these states can actually
survive for cosmological times, as long as the product of the black hole
and the scalar field mass is sufficiently small. Furthermore, we provided strong
evidence in Ref.~\cite{Barranco:2012qs} that these quasi-bound modes are generic.
This conclusion was reached by numerically evolving arbitrary initial data and
Fourier-transforming the solution with respect to time. The spectrum revealed a
clear excitation of the quasi-bound frequencies.

In this paper we use the Green's function representation technique~\cite{Leaver:1986gd} to determine 
the late time behavior of localized scalar field distributions on a 
Schwarzschild background. Within this technique, the Green's function of the problem is decomposed into 
three contributions, corresponding 
to different contours in the complex frequency plane. The contribution we are focusing on in this work 
is the one arising from the residua of the poles of the Green's function, which describe the excitations 
of the quasi-bound states. This allows us to predict, given arbitrary initial data for the scalar field 
on a spacelike hypersurface, the amplitude of each quasi-bound state in the solution by computing a 
simple, one-dimensional integral. Furthermore, we show by comparison with results obtained from numerical 
evolutions that for large times (within the time scales of our simulations) the behavior of the solution is 
well approximated by the corresponding superposition of quasi-bound sates; therefore, the remaining two 
contributions to the Green's function become negligible for those times.
Note that this situation is different than the case of a massless scalar field propagating on a 
Schwarzschild background, where the oscillating part of the solution, described by a superposition of 
quasi-normal modes, is soon taken over by the slower, polynomial tail decay. This difference resides 
in the fact that the quasi-bound modes have a decay rate which is much smaller than the quasi-normal 
modes, so that they dominate the solution for much larger time. 

At this point we would like to 
stress that our definition of ``large" or ``late" times in this article is 
the following: we consider time scales which are very large compared to the Schwarzschild radius of the 
black hole, but which are within the lifetime of the quasi-bound states. In particular, we do not claim 
that for $t\to \infty$ the solution is described by a simple superposition of quasi-bound states since in 
this limit the contribution from the tail part is expected to play an important role. However, for the 
physical scenario we are interested in, it is enough to consider large times which, according to our definition, could 
be larger than the age of the universe as we explained above.

Our results provide the basis for a generalization
to the case of a rotating black hole. Several lines of work have been developed in the context
of scalar and other fields on Kerr background; there are those which consider unstable modes
in the vicinity of the Kerr black hole \cite{Zouros:1979iw, Pani:2012bp}, and even possible radiation
emission due to such instabilities \cite{Yoshino:2012kn, Yoshino:2013ofa}. The continuation of our work, 
however, will focus on a somewhat different direction, namely the scalar field in the Kerr background
in connection with dark matter halos.

The remaining part of this work is organized as follows. In Sec.~\ref{Sec:Green} we briefly review the 
Green's function method and the corresponding decomposition of the 
solution in three different contributions. Next, in
Sec.~\ref{Sec:Toy} we discuss the case of a potential formed by a delta and a step function, as a toy 
model for which the modes can be represented analytically. 
In particular, we determine the quasi-bound frequencies in the limit for which the height of the step is 
very small compared to the amplitude of the delta barrier
and show that they are located very close to the resonant band, resulting in very small decay rates. Then, we 
compute the excitations of
each quasi-bound state, compare the resulting superposition to the results obtained from a numerical 
evolution, and find a very good agreement at late times. 

In Sec.~\ref{Sec:KG} we apply the Green's function technique to the case of a massive scalar field on a Schwarzschild background. 
Here the modes and quasi-bound frequencies are computed semianalytically
via a matching algorithm from which the excitation amplitudes can be 
determined. In Sec.~\ref{Sec:Numerical} we compare the semianalytic results with those obtained from numerical 
evolutions of the Cauchy problem, and as in the case of the toy model, for late times we
find an excellent agreement between the two approaches. Then, in Sec.~\ref{Sec:Discussion} we use our 
semianalytic solution to extract some physical properties of the quasi-bound states that might be
relevant in the astrophysical context, and finally in Sec.~\ref{sec:conclusions} we present our conclusions. 

We also include two appendixes. In the first one we generalize the Green's function technique to time foliations 
different than the one arising from standard
Schwarzschild coordinates. In the second appendix we provide some estimates on the location of the 
quasi-bound frequencies for the toy model example and the 
time after which the solution can be described by a superposition of the corresponding modes.

%%%%%%%%%%%%%%%%%%%%%%%%%%%%%%%%%%%%%%%%%%%%
%%%%%%%%%%%%%%%%%%%%%%%%%%%%%%%%%%%%%%%%%%%%
\section{Green's function representation and spectroscopy}
\label{Sec:Green}
%%%%%%%%%%%%%%%%%%%%%%%%%%%%%%%%%%%%%%%%%%%%
%%%%%%%%%%%%%%%%%%%%%%%%%%%%%%%%%%%%%%%%%%%%

In this section we briefly review the Green's function representation technique and the corresponding
decomposition of the integral kernel. We will concentrate on the solution of $1+1$ dimensional wave problems with time-independent, 
non-negative potentials $V(x)$; see the original paper by Leaver~\cite{Leaver:1986gd} and the review 
articles~\cite{Nollert:1999ji,Kokkotas99a,Berti:2009kk}
for further details. Therefore, we consider the Cauchy problem
\begin{subequations}\label{Eq.Cauchy}
\begin{equation}\label{Eq:Cauchy1}
\frac{\partial^2 \phi}{\partial t^2} - \frac{\partial^2 \phi}{\partial x^2}  + V(x)\phi = 0,
\end{equation}
with $t > 0$, $x\in\Real$, and initial data
\begin{equation}\label{Eq:Cauchy2}
\phi(0,x) = \phi_0(x),\quad 
\frac{\partial \phi}{\partial t}(0,x) = \pi_0(x).
\end{equation}

\end{subequations}
We assume $\phi_0$ and $\pi_0$ to be smooth and compactly supported. In this paper we will be interested only in the late time behavior.

The solution of the problem in Eqs.~(\ref{Eq.Cauchy}) can be represented in terms of an integral kernel $k(t,x,y)$,
\begin{equation}
\phi(t,x) = \int\limits_{-\infty}^\infty \frac{\partial k}{\partial t}(t,x,y)\phi_0(y) dy
 + \int\limits_{-\infty}^\infty k(t,x,y)\pi_0(y) dy,
\label{Eq:IntegralRepresentation}
\end{equation}
where $k(t,x,y)$ can formally be written as
\begin{equation}
k(t,x,y) = \frac{1}{2\pi i}\int\limits_{s=\eta-i\infty}^{\eta+i\infty} e^{st} G(s,x,y) ds,\qquad
\eta > 0,
\label{Eq:Kernel}
\end{equation}
with $G(s,x,y) = G(s,y,x)$ the Green's function,
\begin{equation}
G(s,x,y) := \frac{1}{W(s)}\left\{ \begin{array}{ll} 
 f_-(s,y) f_+(s,x),  & y\leq x \\ 
 f_-(s,x) f_+(s,y), & y > x 
\end{array} \right\}.
\label{Eq:GreenFct}
\end{equation}
Here $f_+(s,x)$ and $f_-(s,x)$ are two nontrivial solutions of the mode equation 
$s^2 f - f'' + V(x) f = 0$ with exponential decay at $x\to +\infty$ and $x\to -\infty$,
respectively, when $\re(s) > 0$, and the prime denotes the derivative with respect to the variable $x$. 
$W(s)$ is their Wronski determinant defined as
\begin{equation}
W(s) := \det\left( \begin{array}{ll}
 f_+(s,x) & f_-(s,x) \\
 f_+'(s,x) & f_-'(s,x)
\end{array} \right).
\label{Eq:WronskiDet}
\end{equation}
Note that $W(s)$ does not depend on $x$, and by definition, it is zero if and only if 
the two solutions $f_+(s,\cdot)$ and $f_-(s,\cdot)$ are proportional to each other.

\begin{figure}[ht]
  \begin{center}
    \includegraphics[angle=0,width=0.4\textwidth,height=!,clip]{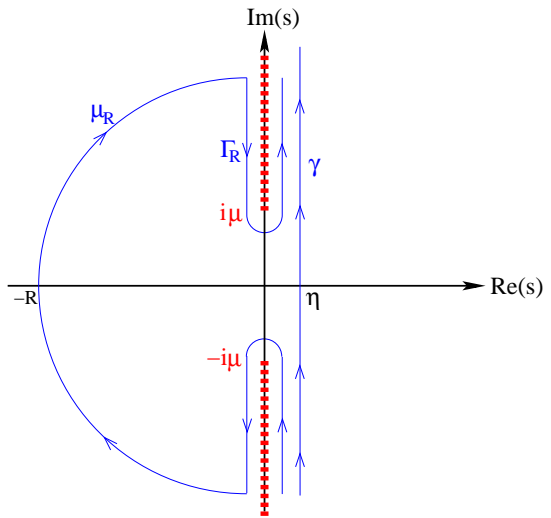}
  \end{center}
  \caption{The branch cut (broken lines) for the analytic function
    $\kappa(s)$ defined in Eq.~(\ref{Eq:kappaDef}) and the deformation of the
    curve $\gamma$ which allows to close the integration contour. }
\label{f:contour}
\end{figure}

A priori, the Green's function is a well-defined, analytic function on the right complex plane $\re(s) > 0$. However, it is interesting to consider its analytic continuation to the left complex plane $\re(s) < 0$ since this offers the possibility to ``close'' the integration contour in Eq.~(\ref{Eq:Kernel}), as illustrated in Fig.~\ref{f:contour}. As a consequence, the kernel typically splits into a sum of three different contributions,
\begin{equation}
k(t,x,y) = k_{\textrm{modes}}(t,x,y) + k_{\textrm{tail}}(t,x,y) + k_{\textrm{hfa}}(t,x,y).
\end{equation}
Here, $k_{\textrm{modes}}$ is a sum over the residua of the poles of the analytic continuation of the Green's function, and each term in the sum is proportional to $e^{s_n t}$. These terms describe the resonances; they oscillate with frequency $\omega_n = \im(s_n)$ and their amplitude decays exponentially in time since $\re(s_n) < 0$. Next, $k_{\textrm{tail}}$ is the contribution that comes from the integration around the branch cuts, and it is usually associated with the tail decay, giving rise to a polynomial decay of the form $t^{-p}$ for some $p > 0$. Finally, $k_{\textrm{hfa}}$ is the contribution from the high-frequency arc (see the curve $\mu_R$ in Fig.~\ref{f:contour}). This contribution is highly dependent on the initial data; however, it vanishes for large enough $t$.

Of course, the structure of the analytic continuation of the Green's function (including the location of the poles and the branch cuts) depends on the potential $V(x)$ and the problem at hand. In the next section, we analyze this structure for the case of a simple toy model, and we show that all the poles are located close to the resonant band $i(-\mu,\mu)$ and give rise to quasi-bound states. Based on this analysis we expect the structure of the analytic continuation to be very similar in the case of a massive scalar field propagating on a Schwarzschild black hole.

In this work we are mainly interested in the contribution from the poles, that is, $k_{\textrm{modes}}$, since in our context this is the part corresponding to the quasi-bound states.
The poles, $s_n$, of the Green's function are determined by the zeros of the Wronskian $W(s)$; notice that at such zeros the functions $f_+(s_n,\cdot)$ 
and $f_-(s_n,\cdot)$ are linearly dependent.
Assuming that $W(s)$ has only simple zeros at $s = s_n$, that is, $W(s) = \alpha_n(s - s_n) + {\cal O}(s - s_n)^2$, with
\begin{equation}
\alpha_n := \left. \frac{d}{ds} W(s) \right|_{s=s_n} \neq 0,
\end{equation}
the mode contribution to the solution from the $n$'th mode $s_n$ is given by
\begin{subequations}\label{Eq:mode.amplitude}
\begin{equation}
\phi_{\textrm{modes},n}(t,x) = A_n e^{s_n t} f_+(s_n,x),
\label{Eq:ModeSolution}
\end{equation}
with amplitude
\begin{eqnarray}
A_n &:=& \left. \left[ \frac{d}{ds} W(s) \right]^{-1} \right|_{s=s_n}\nonumber\\
 &\times& \int\limits_{-\infty}^\infty f_-(s_n,y) [s_n\phi_0(y) + \pi_0(y) ] dy.
\label{Eq:Amplitude}
\end{eqnarray}
\end{subequations}
When the potential $V(x)$ in Eq.~(\ref{Eq:Cauchy1}) is real, the poles come in complex conjugate pairs, $s_{-n} = s_n^*$, and since $W(s)$ and $f_\pm(s,x)$ are analytic 
functions in $s$, the sum $\phi_{\textrm{modes},n}(t,x) + \phi_{\textrm{modes},-n}(t,x)$ is real.

For a generalization of these expressions to more general foliations by spacelike hypersurfaces, see Appendix~\ref{App:A}.

%%%%%%%%%%%%%%%%%%%%%%%%%%%%%%%%%%%%%%%%%%%%
%%%%%%%%%%%%%%%%%%%%%%%%%%%%%%%%%%%%%%%%%%%%
\section{A toy model example}
\label{Sec:Toy}
%%%%%%%%%%%%%%%%%%%%%%%%%%%%%%%%%%%%%%%%%%%%
%%%%%%%%%%%%%%%%%%%%%%%%%%%%%%%%%%%%%%%%%%%%

Before applying the Green's function technique just described to the scalar field problem on a Schwarzschild black hole %, Sec.~\ref{Sec:KG}, 
we consider the following simple toy model potential~\cite{Barranco:2012qs}:
\begin{equation}
V(x) = A\delta(x) + \mu^2\theta(x-a),
\label{Eq:ToyModelPotential}
\end{equation}
consisting of a delta-barrier with positive amplitude $A$ at $x=0$ and a step function at 
$x=a > 0$, where $V(x)$ jumps from zero to its asymptotic value $\mu^2$. (Do not confuse the amplitude $A$ in the potential, with the amplitudes $A_n$ of the quasi-bound modes.) 
As it turns out, this model captures all the rough qualitative features of the scalar field problem.

In order to construct the Green's function for the potential in Eq.~(\ref{Eq:ToyModelPotential}), we consider first the region $x < 0$, where $V(x)=0$, such that
\begin{subequations}
\begin{eqnarray}
f_-(s,x) &=& e^{sx},\\
f_+(s,x) &=& a_+(s) e^{-sx} + b_+(s) e^{sx}.
\end{eqnarray}
\end{subequations}
Notice that $f_-(s,\cdot)$ decays exponentially at $x\to -\infty$ when $\re(s) > 0$, as required. The coefficients $a_+(s)$ and $b_+(s)$ are unknown to this point and will be determined by the
matching conditions. In the region $0 < x < a$, the two solutions have the form
\begin{subequations}
\begin{eqnarray}
f_-(s,x) &=& c_-(s) e^{-sx} + d_-(s) e^{sx},\\
f_+(s,x) &=& c_+(s) e^{-sx} + d_+(s) e^{sx},
\end{eqnarray}
\end{subequations}
with coefficients $c_\pm(s)$, $d_\pm(s)$. Finally, in the region $x > a$, where $V(x) = \mu^2$, we have
\begin{subequations}
\begin{eqnarray}
f_-(s,x) &=& a_-(s) e^{-\kappa(s) x} + b_-(s) e^{\kappa(s)x},\\
f_+(s,x) &=& e^{-\kappa(s) x},
\end{eqnarray}
\end{subequations}
with coefficients $a_-(s)$ and $b_-(s)$; and where we have defined $\kappa(s) := \sqrt{s^2 + \mu^2}$ with the choice for the sign such that $\re(\kappa(s)) > 0$ for $\re(s) > 0$, such that
$f_+(s,\cdot)$ has the required asymptotic behavior at $x\to +\infty$.

Because of the delta-barrier, the matching conditions at $x=0$ consist of the 
continuity of the functions $f_\pm(s,\cdot)$, and the jump condition $f_\pm'(s,0^+) - f_\pm'(s,0^-) =
Af_\pm(s,0)$ for their derivatives. At $x=a$, both $f_\pm(s,\cdot)$ and $f_\pm'(s,\cdot)$ need to be continuous.
These matching conditions yield the following values for the coefficients $c_\pm(s)$ and $d_\pm(s)$:
\begin{subequations}
\begin{eqnarray}
c_-(s) &=& -\frac{A}{2s},
\label{Eq:c-}\\
d_-(s) &=& 1 + \frac{A}{2s},
\label{Eq:d-}\\
c_+(s) &=& \frac{1}{2s}(\kappa + s) e^{-(\kappa-s)a},
\label{Eq:c+}\\
d_+(s) &=& -\frac{1}{2s}(\kappa - s) e^{-(\kappa+s)a}.
\label{Eq:d+}
\end{eqnarray}
\end{subequations}
This information is already sufficient to compute the mode solutions $\phi_{\textrm{mode}}(t,x)$ inside the potential well, $0 < x < a$, assuming that the initial data is supported inside this well.
For this,
we first compute the Wronskian,
\begin{eqnarray}
W(s) &=& 2s\det\left( \begin{array}{ll}
 c_+(s) & c_-(s) \\
 d_+(s) & d_-(s)
\end{array} \right) 
\nonumber\\
 &=& 2s c_+(s) c_-(s)
\left[ \frac{d_-(s)}{c_-(s)} - \frac{d_+(s)}{c_+(s)} \right],
\end{eqnarray}
where we have factored out the non vanishing terms $c_-(s)$ and $c_+(s)$. From this we conclude that the zeros of the determinant are equal to the zeros of the function
\begin{equation}
F(s) := \frac{d_+(s)}{c_+(s)} - \frac{d_-(s)}{c_-(s)}
 = 1 + \frac{2s}{A} - \frac{\kappa-s}{\kappa+s} e^{-2sa}.
\label{Eq:FDef}
\end{equation}
With these observations Eqs.~(\ref{Eq:mode.amplitude}) yield
\begin{eqnarray}
&& \phi_{\textrm{modes},n}(t,x) \nonumber\\
&& = C_n \left[ e^{s_n(t-x)} - \left( 1 + \frac{2s_n}{A} \right) e^{s_n(t+x)} \right],
\label{eq:ModeSoltoy}
\end{eqnarray}
for $0 < x < a$, with the amplitude $C_n$ given by
\begin{eqnarray}
C_n &=& B_n\int\limits_0^a \left[ e^{-s_n y} 
- \left( 1 + \frac{2s_n}{A} \right) e^{s_n y} \right]
\nonumber\\
&\times& \left[Ês_n\phi_0(y) + \pi_0(y) \right] dy,
\label{Eq:CnDef}
\end{eqnarray}
and
\begin{equation}
B_n := -\frac{1}{2s_n}  \left. \left[ \frac{d}{ds} F(s) \right]^{-1} \right|_{s=s_n}.
\label{Eq:BnDef}
\end{equation}
In deriving these equations we have used the fact that, for $0 < x < a$,
\begin{eqnarray}
f_+(s_n,x) &=& c_+(s_n) \left( e^{-sx} + \frac{d_+(s_n)}{c_+(s_n)} e^{sx} \right) \nonumber\\
 &=& c_+(s_n) \left( e^{-sx} + \frac{d_-(s_n)}{c_-(s_n)} e^{sx} \right).
 \end{eqnarray}
Therefore, in order to compute the mode solutions, one first needs to determine the 
zeros of the Wronskian, which coincide with the zeros of the function $F(s)$ defined in
Eq.~(\ref{Eq:FDef}). Next, one needs to determine the quantities $B_n$ from
Eq.~(\ref{Eq:BnDef}), and from this one can compute the amplitude for arbitrary initial
data $\phi_0$ and $\pi_0$ from Eq.~(\ref{Eq:CnDef}). In the following, we compute the
poles and the corresponding amplitude coefficients $B_n$ in the limit $\mu \ll A$.

\subsection{Computation of the poles}

Before computing the zeros of the function $F$, it is important to discuss the analytic 
continuation of the function $\kappa(s) = \sqrt{s^2 + \mu^2}$ for $\re(s) \leq 0$. For
what follows, we define
\begin{equation}
\kappa(s) := \exp\left[\frac{1}{2}\log(s^2 + \mu^2) \right],
\label{Eq:kappaDef}
\end{equation}
which is well-defined for all $s\in\Complex$ except those for which the argument of the 
logarithm is negative or zero. Therefore, $\kappa(s)$ is defined for all $s$ belonging to the set
\begin{equation}
U := \{ s\in \Complex : s\notin \pm i[\mu,\infty) \},
\label{Eq:UDef}
\end{equation}
see Fig.~\ref{f:contour}. This continuation has the property that $\re(\kappa(s)) > 0$ 
for all $s\in U$, implying that the function $f_+(s,\cdot)$ decays exponentially at $x\to
+\infty$ for all such $s$. Note that this is different than what occurs for the quasi-normal modes, where for $\re(s) < 0$ the function $f_+(s,\cdot)$ grows 
exponentially when $x\to +\infty$. It turns out
that for our purposes the definition in Eq.~(\ref{Eq:kappaDef}) provides the correct choice of analytic continuation, since it contains the complex interval 
$i(-\mu,\mu)$ (the resonant band), close to which the quasi-bound modes lie.

\begin{figure}[ht]
  \begin{center}
    \includegraphics[angle=0,width=0.4\textwidth,height=!,clip]{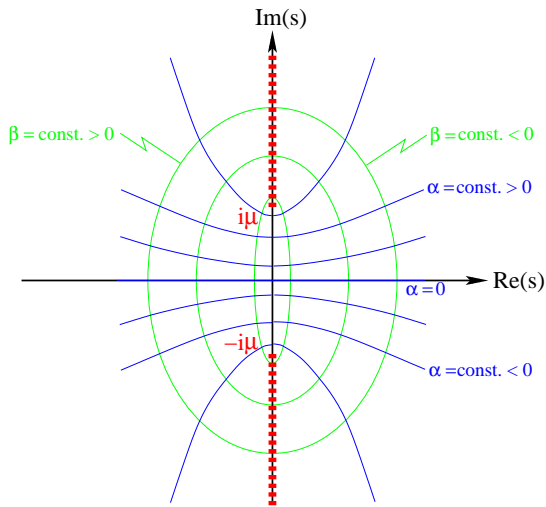}
  \end{center}
  \caption{Visualization of the complex angle $\varphi = \alpha + i\beta$ in the $s$-plane.}
\label{f:newcoords}
\end{figure}

For the following calculations it is convenient to replace $s\in U$ with the complex 
angle $\varphi$ defined by the analytic, one-to-one transformation
\begin{equation}
Q\to U, \varphi \mapsto s:=i\mu\sin(\varphi),
\label{Eq:varphi}
\end{equation}
with $Q:= \{ \varphi\in\Complex : -\pi/2 < \re(\varphi) < \pi/2 \}$. The lines $\re(\varphi) =  \textrm{const.}$ describe hyperbolas in the $s$-plane with focal points $\pm i\mu$, and when
$\re(\varphi)\to
\pm\pi/2$, these hyperbolas approach the branch cuts $\pm i[\mu,\infty)$. The lines $\im(\varphi) =  \textrm{const.}$, in turn, describe semi-ellipsis in the $s$-plane with focal points $\pm i\mu$. In
the limit
$\im(\varphi)\to 0$ these lines approach the complex interval $i[-\mu,\mu]$; see Fig.~\ref{f:newcoords}.

Working with the complex angle $\varphi$ instead of $s$ simplifies the following calculations. For example, $\kappa(s) = \mu\cos(\varphi)$ and $\kappa \pm s = e^{\pm i\varphi}$ have simpler
representations, and the function $F$ defined in Eq.~(\ref{Eq:FDef}) has the form
\begin{equation}
F(\lambda,\varphi) := 1 + i\lambda\sin\varphi - e^{-2i\varphi - 2qi\sin\varphi},
\label{Eq:FDef2}
\end{equation}
where the parameter $q$ is defined as $q := a\mu$. As shown in Ref.~\cite{Barranco:2012qs}, the zeros of $F$ can be determined analytically based on a series expansion in $\lambda := 2\mu/A$. When
$\lambda=0$, the zeros of $F(0,\varphi)$ are given by
\begin{equation}
e^{-2i\varphi - 2qi\sin\varphi} = 1,
\end{equation}
whose solutions $\varphi_n^{(0)}\in [-\pi/2,\pi/2]$ are real and satisfy
\begin{equation}
q\sin(\varphi_n^{(0)}) = n\pi - \varphi_n^{(0)},\qquad n=-N_q,\ldots,N_q.
\label{Eq:phi0Sol}
\end{equation}
They give rise to a finite set of purely imaginary frequencies $s = i\omega_n^{(0)}$ inside the resonant band, with $\omega_n^{(0)} = \mu\sin\varphi_n^{(0)}$.

\begin{figure}[ht]
  \begin{center}
    \includegraphics[angle=0,width=0.35\textwidth,height=!,clip]{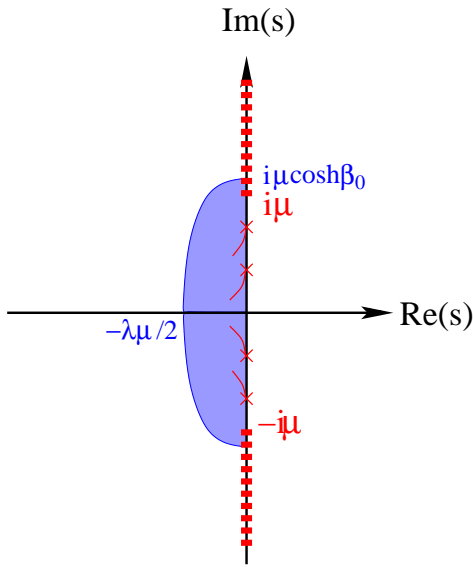}
  \end{center}
  \caption{The location of the poles of the Green's function for small $\lambda > 0$ are
shown by the crosses and the lines emanating from them. There are no
poles outside the shaded region.}
\label{f:poles}
\end{figure}

For small values of $\lambda > 0$ we introduce the quantity $q_n := 2q + 2/\cos(\varphi_n^{(0)})$ , and the zeros of $F(\lambda,\varphi)$ can be written as the following series expansion:
\begin{widetext}
\begin{equation}
\varphi_n(\lambda) = \varphi_n^{(0)} - \frac{\tan(\varphi_n^{(0)})}{q_n} \lambda
 + \tan(\varphi_n^{(0)})\left( \frac{1}{q_n^2} + \frac{q}{q_n^3}\tan^2(\varphi_n^{(0)})
  + \frac{i}{2}\frac{\sin(\varphi_n^{(0)})}{q_n} \right) \lambda^2 + {\cal O}(\lambda^3),
\label{Eq:phinExpansion}
\end{equation}
which yields
\begin{equation}
s_n(\lambda) = 
 i\omega_n^{(0)}\left[ 1 - \frac{\lambda}{q_n} + \frac{\sigma_n}{q_n^2}\lambda^2
 + \frac{i}{2}\frac{\sin(\varphi_n^{(0)})}{q_n}\lambda^2
 +  {\cal O}(\lambda^3) \right],
\label{Eq:snExpansion}
\end{equation}
\end{widetext}
with $\sigma_n := 1 - q_n^{-1}\sin^2(\varphi_n^{(0)})/\cos^3(\varphi_n^{(0)})$. This expansion shows that, when $\lambda > 0$, the poles, which for $\lambda = 0$ lie all on the imaginary interval
$i(-\mu,\mu)$, wander to the left half-plane, with the real part scaling like $\lambda^2$ for small $\lambda$; see Fig.~\ref{f:poles}.

It follows from the estimate Eq.~(\ref{Eq:FEstimate}) in Appendix~\ref{App:B} that $|F(\lambda,\varphi)| \geq e^\beta(2\sinh\beta - \lambda)$ for all $\varphi = \alpha +  i\beta$ with $\beta > 0$,
implying that there are no poles in the left-half plane which lie outside the semi-ellipse $\im(\varphi) = \beta_0 := \arcsinh(\lambda/2)$. Therefore, when $\lambda > 0$ is small, all poles must lie
inside a small region to the left of the resonant band $i(-\mu,\mu)$; see Fig.~\ref{f:poles} again. This is different from the quasi-normal mode case, where there are poles arbitrarily far away from
the
origin.

\subsection{Computation of the coefficients $B_n$}

In order to compute the coefficients $B_n$ defined in Eq.~(\ref{Eq:BnDef}), we first notice that
\begin{eqnarray}
&& \left[ \frac{d}{ds} F(s) \right]_{s=s_n}
 = \left[ \frac{\partial F(\lambda,\varphi)}{\partial\varphi} 
\left( \frac{ds}{d\varphi} \right)^{-1} \right]_{s=s_n}
\\
 && = \frac{1}{\mu}\left[ 2\left(1 + i\lambda\sin\varphi_n(\lambda) \right)
 \left( q +\frac{1}{\cos\varphi_n(\lambda)} \right) + \lambda \right],
\nonumber
 \end{eqnarray}
from which
\begin{widetext}
\begin{equation}
B_n^{(\lambda)} = \frac{i}{2\sin\varphi_n(\lambda)}
\left[ 2\left(1 + i\lambda\sin\varphi_n(\lambda) \right)
 \left( q +\frac{1}{\cos\varphi_n(\lambda)} \right) + \lambda \right]^{-1}.
\end{equation}
Here, we can substitute the expansion~(\ref{Eq:phinExpansion}) for $\varphi_n(\lambda)$, which yields
\begin{subequations}
\begin{eqnarray}
\sin\varphi_n(\lambda) &=&  \sin\varphi_n^{(0)}
\left[ 1 - \frac{1}{q_n}\lambda + \frac{\sigma_n}{q_n^2}\lambda^2
 + \frac{i}{2}\frac{\sin\varphi_n^{(0)}}{q_n}\lambda^2
 +  {\cal O}(\lambda^3) \right],
\\
\cos\varphi_n(\lambda) &=&  \cos\varphi_n^{(0)}
\left[ 1 + \frac{\tan^2\varphi_n^{(0)}}{q_n}\lambda 
 - \tan^2\varphi_n^{(0)}\left( \frac{1}{q_n^2} 
  + \frac{1}{q_n^3}\frac{q + \cos\varphi_n^{(0)}}{\cos^2\varphi_n^{(0)}}
 + \frac{i}{2}\frac{\sin\varphi_n^{(0)}}{q_n} \right)\lambda^2
 +  {\cal O}(\lambda^3) \right].
\end{eqnarray}
\end{subequations}
\end{widetext}

\subsection{The case $\lambda = 0$}

It is instructive to discuss in more detail the limit $\lambda\to 0$, corresponding to a delta-barrier with infinite amplitude, i.e. a hard wall. 
In this case one expects \emph{normal modes}, i.e. purely
imaginary poles for the Green's function. Indeed, when $\lambda=0$, we saw that the zeros of the Wronskian are purely imaginary, $s_n^{(0)} = i\mu\sin(\varphi_n^{(0)})$, where $\varphi_n^{(0)}\in
(-\pi/2,\pi/2)$ are the solutions of Eq.~(\ref{Eq:phi0Sol}).

In the limit $\lambda\to 0$, the amplitude coefficients $B_n$ reduce to
\begin{equation}
B_n^{(0)} = \frac{i}{4\omega_n^{(0)}} 
\frac{\sqrt{\mu^2-(\omega_n^{(0)})^2}}{1 + a\sqrt{\mu^2-(\omega_n^{(0)})^2}},
\end{equation}
where we recall that $\omega_n^{(0)} = \mu\sin\varphi_n^{(0)}$. Taking into account 
that $2s_n/A = i\lambda\sin\varphi_n(\lambda)$, the expression~(\ref{Eq:CnDef}) for 
the coefficients $C_n$ reduces to
\begin{eqnarray}
C_n^{(0)} &=& \frac{1}{2\omega_n^{(0)}} 
\frac{\sqrt{\mu^2-(\omega_n^{(0)})^2}}{1 + a\sqrt{\mu^2-(\omega_n^{(0)})^2}}
\nonumber\\
&\times&\int\limits_0^a \sin(\omega_n^{(0)} y)( i\omega_n^{(0)} \phi_0(y) + \pi_0(y) ) dy.
%\nonumber
\end{eqnarray}
Therefore, the mode solution in this limit is
\begin{equation}
\phi_{\textrm{modes}}(t,x) = 
 4\im\left[ \sum\limits_{n=1}^{N_q} C_n^{(0)} e^{i\omega_n^{(0)} t}
  \sin(\omega_n^{(0)} x) \right],
\end{equation}
for $0 < x < a$, and as expected, it is a sum over purely oscillating factors.

\subsection{Numerical results}\label{subsec:numerical}

To complement our previous analytic results, we solve numerically the Cauchy problem in Eqs.~(\ref{Eq.Cauchy}) and~(\ref{Eq:ToyModelPotential}). First, we define a new set of
variables, $\pi=\partial_{t}\phi$ and $D = \partial_{x}\phi$, to convert Eq.~(\ref{Eq:Cauchy1}) into a first-order system of linear differential equations. Then, we apply the method of lines with a
third order iterated Crank-Nicholson integrator, and use second order finite differences for the spatial discretization. We work on the spatial domain 
$[0,x_{\textrm{max}}]$, with $x_{\textrm{max}}$ larger than $a$. The
boundary condition imposed at the origin, $x=0$, is $\partial_{x}\phi_{R} = \partial_{t}\phi_{R} + A\phi_{R}$, where the labels refer to the left and right of $x = 0$. This condition comes from the
behavior
of the function dictated by the delta-potential. To see this, notice that due to the continuity at $x = 0$, $\phi$ must satisfy that $\phi_{R} = \phi_{L}$, and $\partial_{x}\phi_{R} =
\partial_{x}\phi_{L}+A\phi_{L}$. If we 
impose that the flux is purely outgoing at $x < 0$, we have an extra condition, namely $\partial_{t}\phi_{L} = \partial_{x}\phi_{L}$. After eliminating $\partial_{t}\phi_{L}$ from these equations, we
arrive to the desired condition. 
At the right boundary $x = x_{\textrm{max}}$ we use an outgoing Sommerfeld-like condition. 

\begin{figure}[ht]
\begin{center}
\includegraphics[angle=0,width=0.5\textwidth,height=!,clip]{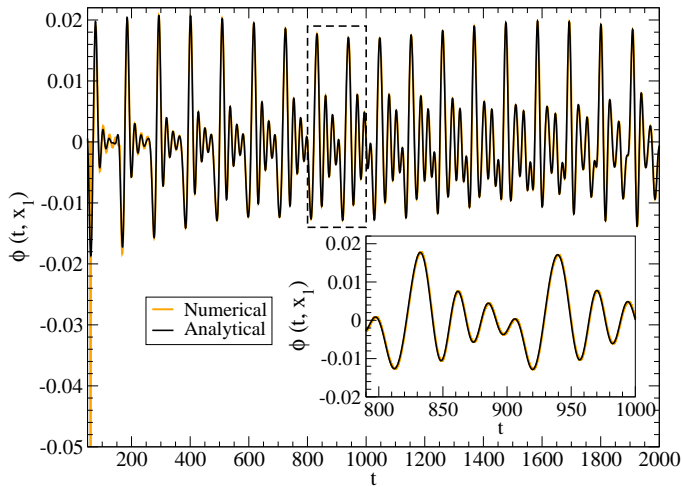}
\end{center}
\caption{For a fixed observer at $x_1=45$, we plot the wave function obtained as the contribution of the first five quasi-bound modes (twice the real part of the sum over $n = 1,2,\ldots,N_q=5$ of the modes defined in Eq.~(\ref{eq:ModeSoltoy})) and the numerical solution. The parameters for the potential used in this example
are: $A=6$, $\mu=0.3$, $a=50$, and the parameters for the initial data are: $\tilde{\phi}_0 = 0.1$, $x_0 = 15$, $\sigma=1$. The relative numerical error is estimated to be less than $1\%$.
}
\label{fig:comp_m03A6a50z} 
\end{figure}

We find a very good agreement between the numerical solution and the function given
by the sum of modes defined in Eq.~(\ref{eq:ModeSoltoy}). In Fig.~\ref{fig:comp_m03A6a50z}, we present the numerical solution and the function obtained from the sum over the contribution of the modes. This example corresponds to a Gaussian initial data of the form $\phi_{0}(x) = \tilde\phi_{0}e^{-(x-x_0)^2/\sigma^2}$ and $\pi_{0}(x) = D_{0}(x)$. After an initial transient period, the contribution from the quasi-bound modes dominate and the solution can be accurately represented by the sum of these modes. 

The fact that at large times the solution can be accurately described solely in terms of the quasi-bound modes is remarkable, since as discussed in Sec.~\ref{Sec:Green} the solution is really a sum
over three contributions. In order to get some understanding about this fact, we analyze in Appendix~\ref{App:B} the contribution from the high-frequency arc and show that it vanishes identically for
times larger than $2a$, corresponding to two crossing times of the potential well. Interestingly, it seems that the tail contribution is also negligible, at least for the time scales of our
simulations. We will find a similar behavior in the next two sections when discussing the Klein-Gordon equation on a Schwarzschild black hole.

\begin{figure}[ht]
\begin{center}
\includegraphics[angle=0,width=0.5\textwidth,height=!,clip]{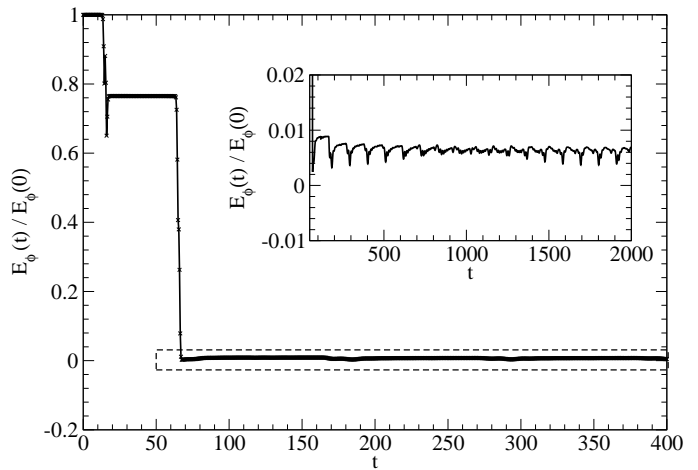}
\end{center}
\caption{After some transient period in which the field bounces between the potential barriers, the
energy drops from its initial value and settles down around a fixed value. This late 
contribution is given by the sum of the modes, Eq.~(\ref{eq:ModeSoltoy}).}
\label{fig:emodes} 
\end{figure}

We also calculate the energy of the field inside the well using the integral $E_{\phi}(t)=\frac{1}{2}\int_{0}^{a}(\pi(t,x)^2+D(t,x)^2  )\,dx$. In Fig.~\ref{fig:emodes} 
we show the evolution of the energy normalized
by the energy of the initial configuration. At late times, when the modes contribution is dominant, the remaining energy for this toy model is about $0.6\%$ of its initial value.

%%%%%%%%%%%%%%%%%%%%%%%%%%%%%%%%%%%%%%%%%%%%
%%%%%%%%%%%%%%%%%%%%%%%%%%%%%%%%%%%%%%%%%%%%
\section{Quasi-bound states and spectroscopy for the Klein-Gordon equation}
\label{Sec:KG}
%%%%%%%%%%%%%%%%%%%%%%%%%%%%%%%%%%%%%%%%%%%%
%%%%%%%%%%%%%%%%%%%%%%%%%%%%%%%%%%%%%%%%%%%%

The evolution of a scalar field with fixed angular momentum number, $\ell$, propagating on a Schwarzschild background spacetime is described by the Cauchy problem in
Eqs.~(\ref{Eq.Cauchy}), where the potential is given now by
\begin{equation}
V(x) = \left( 1 - \frac{2M}{r} \right)
\left( \frac{\ell(\ell+1)}{r^2} + \frac{2M}{r^3} + \mu^2 \right).
\label{Eq:Veff}
\end{equation}
Here $x := r_* = r + 2M\log(r/2M-1)$ is the tortoise radial coordinate, with $r > 2M$; please see Ref.~\cite{Barranco:2012qs} for further details concerning the notation in this section. The physical field $\Phi$ is related to the field $\phi$ satisfying Eq.~(\ref{Eq:Cauchy1}) by
$$
\Phi(t,r,\vartheta,\varphi) = \frac{1}{r}\phi(t,x) Y^\ell(\vartheta,\varphi)
$$
with $Y^\ell$ a spherical harmonic function with associated angular momentum number $\ell$.

Assuming again that all zeros of the Wronskian are simple, we apply the formulae in Eqs.~(\ref{Eq:mode.amplitude}) to the Cauchy problem with potential given in Eq.~(\ref{Eq:Veff}) . The difficulty now is that the functions $f_\pm(s,x)$ and their Wronskian cannot be computed in an explicit simple form, as it was possible in the previous section. As explained in more detail below, the functions $f_\pm(s,x)$ belong to the confluent Heun class, and it is not known to the authors whether or not there are simple expressions for the Wronskian of such functions. Approximations which are valid in the small $M\mu$ limit have been worked out in Ref.~\cite{Detweiler:1980uk}. However, in this work we will not make use of these approximation; rather, we will construct the functions $f_\pm(s,x)$ numerically via a matching procedure.

\subsection{Exact mode solutions}
 
For practical reasons it is convenient to define the dimensionless variables
\begin{equation}
z := \frac{r}{2M} - 1,\qquad
\varepsilon := M\mu .
\end{equation}
Introducing the ansatz $\phi(t,r_*) = e^{st} f(z)$ into Eqs.~(\ref{Eq:Cauchy1}) and~(\ref{Eq:Veff}), we obtain in terms of these new variables
\begin{eqnarray}
&& -\left( \frac{z}{z+1}\frac{d}{dz} \right)^2 f
+ \left[ \frac{\ell(\ell+1)z}{(z+1)^3}
+ \frac{z}{(z+1)^4} - \frac{4\varepsilon^2}{z+1} \right] f
\nonumber\\
&& = -\Omega(s)^2 f.
\label{Eq:ModeEq}
\end{eqnarray}
Here we have defined
\begin{equation}
\Omega(s) := 2\varepsilon\sqrt{1 + \frac{s^2}{\mu^2}}
 = 2\varepsilon\frac{\kappa(s)}{\mu},
\label{Eq:OmegaDef}
\end{equation}
where the function $\kappa(s) = \sqrt{s^2 + \mu^2}$ and its branch points are given in Eq.~(\ref{Eq:kappaDef}). Notice that accordingly, $\Omega(s)$ has positive real part 
for all $s$ in the domain $U$ defined in Eq.~(\ref{Eq:UDef}).

Eq.~(\ref{Eq:ModeEq}) has an irregular singular point at $z = \infty$, and two regular singular points at $z=0$ and $z=-1$, respectively. As $z\to\infty$, the solutions
behave as $e^{\pm\Omega(s) z}$; here we are interested in the one with the minus sign, corresponding to an exponential decay. 
At $z=0$ the characteristic exponents of the regular singular point are $\pm 2Ms$,
meaning the solution behaves as $z^{\pm 2Ms}$ close to $z=0$; the plus sign corresponds to the solution we want to consider which is purely outgoing (that is, moving towards the horizon). At $z=-1$ there is a degenerated characteristic exponent equal
to one. 

An exact solution to the Eq.~(\ref{Eq:ModeEq}) satisfying the required conditions at $z=0$ is given by
\begin{equation}
f_-(s,z) = (z+1) z^{2Ms} e^{-\Omega(s) z}
 \mbox{HeunC}(\alpha,\beta,\gamma,\delta,\eta, -z),
\label{Eq:fmHeun}
\end{equation}
with $\mbox{HeunC}$ the confluent Heun function as defined in MAPLE, the parameters being
\begin{subequations}
\begin{eqnarray}
\alpha &:=& 2\Omega(s),\\
\beta &:=& 4Ms,\\
\gamma &:=& 0,\\
\delta &:=& \Omega(s)^2 + (2Ms)^2,\\
\eta &:=& -\delta - \ell(\ell+1).
\end{eqnarray}
\end{subequations}
Since by definition $\mbox{HeunC}$ is regular in the vicinity of $z=0$, and equal to one at $z=0$, expression~(\ref{Eq:fmHeun}) gives a closed-form representation for the left solution of our problem.

The quasi-bound states correspond to those frequencies for which the left solution, $f_-(s,z)$, decays exponentially when $z\to +\infty$. In view of the closed-form 
expression, Eq.~(\ref{Eq:fmHeun}), this
means that $\mbox{HeunC}$ should not grow faster than polynomially when $z\to\infty$. Unfortunately, the authors are not aware of known asymptotic expansions for the confluent Heun function
$\mbox{HeunC}$ when $z\to \infty$.

\subsection{Hydrogen limit}

For our applications to dark matter halos discussed below, we are interested in small values of $\varepsilon = M\mu > 0$. In this case, the potential $V(r)$ develops a 
well whose minimum is located at $r_{\textrm{min}} \simeq \ell(\ell+1)M/\varepsilon^2$. Therefore, focusing on effects that occur near or to the right of the minimum of the potential well, it is
appropriate to replace $z$ with the
rescaled variable $\zeta := \varepsilon^2 z$. Accordingly, we can replace $\Omega(s)$ with $k(s) := \varepsilon^{-2}\Omega(s)$. When written in terms of $\zeta$ and $k(s)$, the mode
equation~(\ref{Eq:ModeEq}) reduces to
\begin{equation}
-\frac{d^2}{d\zeta^2} f + \left[ \frac{\ell(\ell+1)}{\zeta^2} - \frac{4}{\zeta} \right] f
 = -(k^{(0)}(s))^2 f,
\label{Eq:ModeEqHL}
\end{equation}
when taking the pointwise limit $\varepsilon\to 0$ and assuming that $k(s)\to k^{(0)}(s)$ converges to a finite value in this limit. This is the eigenvalue problem encountered in the discussion of the
hydrogen atom. Therefore, in the limit $\varepsilon \to 0$, the frequencies are determined by the Balmer spectrum for $k^{(0)}(s)$:
\begin{equation}
k_n^{(0)} = \frac{2}{n},\qquad n = \ell+1,\ell+2,\ldots ,
\label{Eq:Balmer}
\end{equation}
with the corresponding eigenfunctions
\begin{equation}
f_n^{(0)}(\zeta) = \zeta^\ell e^{-k_n^{(0)}\zeta}\sum\limits_{j=0}^{n-\ell}
\frac{(\ell + 1 - n)_j}{(2(\ell+1))_j}\frac{\zeta^j}{j!}.
\end{equation}
Here $(a)_0 := 1$, and $(a)_j := a(a+1)\cdot\cdot\cdot(a+j-1)$. In view of 
Eq.~(\ref{Eq:OmegaDef}), and taking $k(s) = \varepsilon^{-2}\Omega(s)$, Eq.~(\ref{Eq:Balmer}) implies 
the following asymptotic expressions for $s$ (compare with Eq.~(49) in Ref.~\cite{Rosa:2011my}):
\begin{equation}
s_n^{(0)} = \pm i\mu\sqrt{ 1 - \frac{\varepsilon^2}{n^2}},\qquad n = \ell+1,\ell+2,\ldots,
\label{Eq:sn0}
\end{equation}
corresponding to a purely imaginary spectrum. This is to be expected, since our considerations here completely neglect effects near the horizon, and as a consequence of our rescaling, the potential is
replaced by the effective potential of the hydrogen problem, which diverges at $\zeta\to 0$. Therefore, the effect of our zeroth order approximation is to replace the potential barrier by an infinite
barrier. Consequently, one finds normal modes instead of quasi-bound states. This is similar to what occurred in the limit $\lambda\to 0$ of the toy model problem 
discussed in the previous section.

As it turns out the zeroth order expression for the frequencies given in Eq.~(\ref{Eq:sn0}) yields already a rather good approximation for the imaginary part of $s_n$, as long as $M\mu\ll 1$. When
$\varepsilon$ is nonzero but small, one still expects the solution $f$ to be described by Eq.~(\ref{Eq:ModeEqHL}) for large enough $z$. The solution of this equation which decays as $z\to\infty$ is
given by~\cite{Detweiler:1980uk}
\begin{equation}
f_+(s,z) = e^{-\xi/2}\xi^{\ell+1} U(\ell+1-\nu,2(\ell+1),\xi),
\label{Eq:fpHydrogen}
\end{equation}
where $\xi := 2\Omega(s)z$, $\nu = 2/k(s)$, and $U(a,b,\xi)$ denotes the confluent hypergeometric function as defined in Ref.~\cite{abramowitz2012handbook}.

\subsection{Shooting to a matching point}

Based on the exact expression for $f_-$ given in Eq.~(\ref{Eq:fmHeun}), and the approximate expression for $f_+$ given in Eq.~(\ref{Eq:fpHydrogen}), we determine by a shooting algorithm the 
complex frequencies, $s_n$, corresponding to the quasi-bound modes. Starting from a given point $r$ in the interval $2M < r < 4M$, and using the known convergent 
series expansion of $\mbox{HeunC}(\alpha,\beta,\gamma,\delta,\eta,-z)$ at $z=0$ (with convergence radius one, see Ref.~\cite{ronveaux1995heun}), this algorithm first numerically integrates the mode
equation~(\ref{Eq:ModeEq}) in
order to extend the function $f_-(s,z)$ to some intermediate point $z_1$. (Typically, we choose $z_1 = \ell(\ell+1)M/\varepsilon^2$ to be located close to the minimum of the potential well.) For the
numerical integration we use a fourth-order Runge-Kutta scheme. Then, the solution $f_+(s,z)$ is extended to $z_1$ in a similar fashion, starting from a large value of $z$ where the
approximation in Eq.~(\ref{Eq:fpHydrogen}) can be used. Then, the Wronski determinant, $W(s)$, defined in 
Eq.~(\ref{Eq:WronskiDet}) is computed at $z_1$. The last ingredient of the algorithm consists of a Newton iteration scheme whose purpose is to find the roots of the Wronskian $W(s)$.

In order to illustrate the effectiveness of the method we use similar initial data as the one used in Refs.~\cite{Barranco:2012qs,Burt:2011pv}. The radial part of the initial data is given by
\begin{equation}
\phi(0,r) = r\times\begin{cases}
  N (r-R_1)^4(r-R_2)^4 & \text{for $R_1 \leq r \leq R_2$} \\
  0 & \text{otherwise}
\end{cases}, \label{eq:id}
\end{equation}
with the normalization $N=[2/(R_1-R_2)]^8$.  The free parameters $R_1$ and $R_2$ allow to set different ``locations'' and ``sizes'' for the scalar field distributions at $t=0$.  The initial value of
the time derivative is chosen as $\dot{\phi}|_{t=0}=\pi_0(r)=0$. 

Next, we show the results for two evolutions corresponding to $M\mu=0.30$ and $M\mu=0.20$, respectively. In both cases $\ell = 1$, and the initial data is supported inside the interval $[R_1,R_2] =
[4M,8M]$, see Eq.~(\ref{eq:id}). The first $9$ quasi-bound frequencies $s_n$ are shown in Table~\ref{table1}.
\begin{table*}[ht]
\caption{Frequencies, $s_n$, for the case $\ell = 1$ and different values of parameters
  $M\mu$ and $n$.} \label{table1}
\begin{tabular}{ lr|rr l|  rr  cc  }
\hline \hline
$n$&&&$M\mu=0.30$&&&&$M\mu=0.20$& \\
&& $\re(Ms_n)$\hfil\hfil && $M\mu-\im(Ms_n)$ \hfil\hfil&&$\re(Ms_n)$\hfil\hfil && $M\mu-\im(Ms_n)$ \hfil \hfil\\
\hline
1  && $-9.4556\times 10^{-6}$ && $3.8076\times 10^{-3}$ && $-4.060\times 10^{-8}$ && $1.0473\times 10^{-3}$ \\
%\hline
2  && $-3.6585\times 10^{-6}$ && $1.6648\times 10^{-3}$ && $-1.473\times 10^{-8}$ && $4.6253\times 10^{-4}$ \\
%\hline
3  && $-1.6491\times 10^{-6}$ && $9.2004\times 10^{-4}$ && $-6.582\times 10^{-9}$ && $2.5845\times 10^{-4}$ \\
%\hline
4  && $-8.6199\times 10^{-7}$ && $5.8091\times 10^{-4}$ && $-3.446\times 10^{-9}$ && $1.6456\times 10^{-4}$ \\
%\hline
5  && $-5.0203\times 10^{-7}$ && $3.9934\times 10^{-4}$ && $-2.014\times 10^{-9}$ && $1.1384\times 10^{-4}$ \\
%\hline
6  && $-3.1649\times 10^{-7}$ && $2.9111\times 10^{-4}$ && $-1.275\times 10^{-9}$ && $8.3395\times 10^{-5}$ \\
%\hline
7  && $-2.1180\times 10^{-7}$ && $2.2152\times 10^{-4}$ && $-8.564\times 10^{-10}$ && $6.3700\times 10^{-5}$ \\
%\hline
8  && $-1.4836\times 10^{-7}$ && $1.7418\times 10^{-4}$ && $-6.021\times 10^{-10}$ && $5.0238\times 10^{-5}$ \\
%\hline
9  && $-1.1332\times 10^{-7}$ && $1.4032\times 10^{-4}$ && $-4.829\times 10^{-10}$ && $4.0487\times 10^{-5}$ \\
\hline \hline
\end{tabular}
\end{table*}

After determining the quasi-bound frequencies, we compute the amplitude each of them is excited from the given initial data. This is performed based on Eq.~(\ref{Eq:AmplitudeBis}) of
Appendix~\ref{App:A}, assuming the initial data is specified on an ingoing Eddington-Finkelstein slice. The corresponding amplitudes $A_n$ for the first 9 modes are shown in Table~\ref{table2}. The
number of modes that are excited is infinite in principle, but as can be seen from Table~\ref{table2}, the amplitudes decrease rapidly for large $n$. More specifically, we have found that for the
particular initial data chosen, the first five modes are relevant, while the sixth one starts to contribute in a negligible way. Of course, the number of modes requires to accurately describing the
modal part of the solution may change depending on the initial data.
\begin{table*}[ht]
\caption{Amplitudes, $A_n$, for the case $\ell = 1$ and different values of parameters
  $M\mu$ and $n$.} \label{table2}
\begin{tabular}{ lr|rr l|  rr  cc  }
\hline \hline
$n$&&&$M\mu=0.30$&&&&$M\mu=0.20$& \\
&& $\re(A_n)$\hfil\hfil && $\im(A_n)$ \hfil\hfil&&$\re(A_n)$\hfil\hfil && $\im(A_n)$ \hfil \hfil\\
\hline
1  && $6.343\times 10^{-4}$ && $5.540\times 10^{-4}$  && $5.87\times 10^{-6}$ && $2.71\times 10^{-6}$ \\
%\hline
2  && $2.404\times 10^{-4}$ && $2.092\times 10^{-4}$  && $2.11\times 10^{-6}$ && $9.87\times 10^{-7}$ \\
%\hline
3  && $1.075\times 10^{-4}$ && $9.357\times 10^{-5}$  && $9.42\times 10^{-7}$ && $4.41\times 10^{-7}$ \\
%\hline
4  && $5.602\times 10^{-5}$ && $4.874\times 10^{-5}$  && $4.93\times 10^{-7}$ && $2.31\times 10^{-7}$ \\
%\hline
5  && $3.256\times 10^{-5}$ && $2.834\times 10^{-5}$  && $2.88\times 10^{-7}$ && $1.35\times 10^{-7}$ \\
%\hline
6  && $2.050\times 10^{-5}$ && $1.785\times 10^{-5}$  && $1.82\times 10^{-7}$ && $8.55\times 10^{-8}$ \\
%\hline
7  && $1.371\times 10^{-5}$ && $1.194\times 10^{-5}$  && $1.22\times 10^{-7}$ && $5.74\times 10^{-8}$ \\
%\hline
8  && $9.598\times 10^{-6}$ && $8.360\times 10^{-6}$  && $8.60\times 10^{-8}$ && $4.04\times 10^{-8}$ \\
%\hline
9  && $7.321\times 10^{-6}$ && $6.392\times 10^{-6}$  && $6.90\times 10^{-8}$ && $3.24\times 10^{-8}$ \\
\hline \hline
\end{tabular}
\end{table*}

\begin{figure*}[ht]
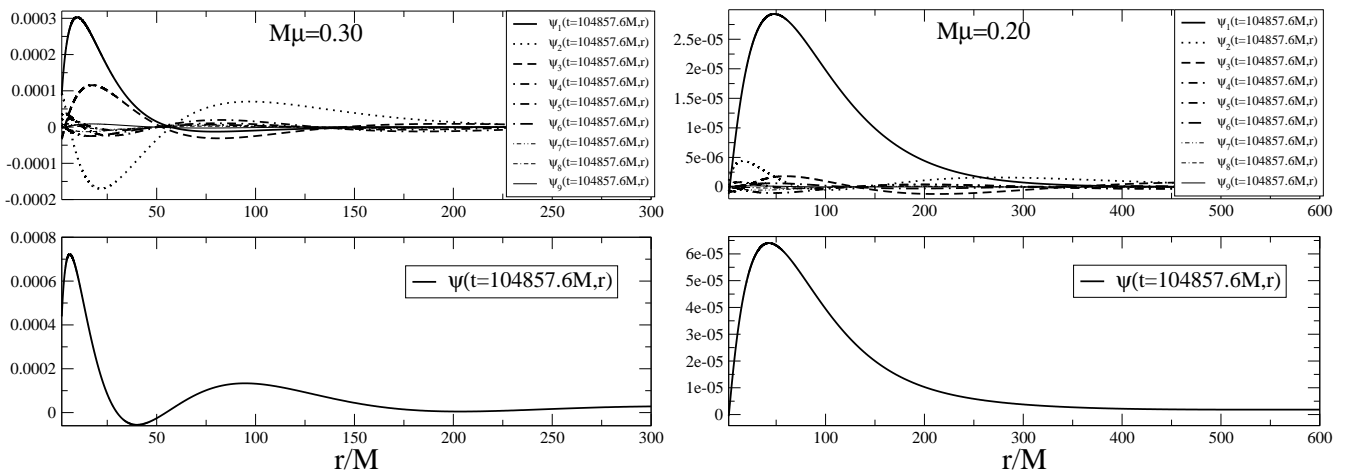

\begin{center}
\includegraphics[angle=0,width=0.49\textwidth,height=!,clip]{mu30theo.eps}
\includegraphics[angle=0,width=0.49\textwidth,height=!,clip]{mu20theo.eps}
\end{center}
\caption{Quasi-bound modes at $t=104857.6M$ for initial data of the form given in Eq.~(\ref{eq:id}), with $R_1 = 4M$ and $R_2 = 8M$. In this case we have fixed $M\mu=0.3$ 
(left), and $M\mu=0.2$ (right). {\it Top panel:} Individual contribution of each excited mode, $\psi_n(t=104857.6M,r)=\phi_n(t=104857.6M,r)/r$, given the initial data. {\it 
Bottom panel:} Sum over the first nine modes.
}
\label{Fig:analytical} 
\end{figure*}

The top panels of Fig.~\ref{Fig:analytical} show the final output for the excited modes,
\begin{equation}
\psi_{n}(t,r) := \re\left( A_n e^{s_n t} f_+(s_n,r)/r \right)\,,
\label{Eq:psi}
\end{equation}
where the values of $A_n$ and $s_n$ are the ones given in Tables~\ref{table1} and \ref{table2}.
The sum of these modes, $\psi(t,r)=\sum_n \psi_n(t,r)$ ($n=1,\ldots, 9$ in our examples) are shown in the lower
panels of Fig.~\ref{Fig:analytical}. In the next section, these solutions will be compared to the numerical
evolutions obtained from the same initial data; see Figs.~\ref{Fig:Comp1} and~\ref{Fig:Comp2} below.

%%%%%%%%%%%%%%%%%%%%%%%%%%%%%%%%%%%%%%%%%%%%
%%%%%%%%%%%%%%%%%%%%%%%%%%%%%%%%%%%%%%%%%%%%
%\section{Dynamical evolution and numerical experiments}
\section{Dynamical evolution and comparison}
\label{Sec:Numerical}
%%%%%%%%%%%%%%%%%%%%%%%%%%%%%%%%%%%%%%%%%%%%
%%%%%%%%%%%%%%%%%%%%%%%%%%%%%%%%%%%%%%%%%%%%

In this section we present some numerical evolutions, with the purpose of 
extending previous work, as well as comparing  with the semianalytical results of
Sec.~\ref{Sec:KG}. We will corroborate that, 
after some transient initial period (and for the time scales reached with our numerical evolutions),
the scalar field can be accurately described by a superposition of quasi-bound modes alone. Furthermore, 
the amplitude of each excited mode can be computed using the recipe presented in the previous section; see for instance Figs.~\ref{Fig:Comp1} and~\ref{Fig:Comp2}, where
the agreement between the numerical evolutions and the semianalytical estimations is remarkable.
As mentioned above, only the first few modes contribute significantly 
for the particular initial data considered in this paper.

\subsection{Dynamical evolutions}

Evolutions of quasi-bound modes on a Schwarzschild background were performed
in Ref.~\cite{Burt:2011pv}. In that case, initial conditions were specifically
chosen in order to excite mainly individual quasi-bound modes.
To our knowledge, long lasting evolutions for more generic
initial data, showing the excitation of a combination of quasi-bound states at
late times, were first performed in Ref.~\cite{Barranco:2012qs} and later also in~\cite{Witek:2012tr}.
In both cases a generic feature is observed:
the excitation of quasi-bound modes, which can be determined (from a Fourier
analysis) by a perfect match between the frequencies in the evolution and
those of the  quasi-bound modes.

We begin this section by extending the study presented in~\cite{Barranco:2012qs} regarding the evolution of arbitrary initial data, considering now a 
larger number of cases. While the initial configurations evolved in previous works were quite wide (from $100M$ to $2000M$) as compared
to the black hole size, here we consider initial data as thin as $2M$. 

Starting from the initial configurations described in the previous section, we
proceed to solve the dynamic equations to obtain the time evolution. The
system is solved numerically with the code used in our previous work,
with the addition of (fixed) mesh refinement
based on the method presented in~\cite{Lehner:2005vc}. This addition
resolved the appearance of noise originated at the right boundary when
performing very large evolutions, as reported in~\cite{Barranco:2012qs}, by
allowing the use of a much larger spatial domain. Aside from
this addition, the original code, of which a  detailed description is
given in~\cite{Megevand:2007uy}, was unchanged. To give a brief description, the code uses second order
finite differences in space and evolves in time using a method of
lines with a third order Runge-Kutta integrator. The use of horizon
penetrating coordinates allows one to use free left boundary conditions (inside
the horizon), while the right boundary is set far away and all incoming modes
are set to zero there.

For the results presented here we set a spatial domain $[r_{\rm min},
  r_{\max}]$ with $r_{\rm min}=1M$ and $r_{\rm max}$ up to about $420,000M$, always
satisfying $r_{\rm max}>T+1200M$, where $T$ is the total evolution time. 
This choice ensures that the region $[r_{\rm min},1200M]$, where we analyze
the results, is not affected by
unphysical signals originated at the right boundary.
A total of 8 grids were used, each one doubling in resolution towards the left, with the highest resolution
being  $\Delta r=0.1M$. In all cases the finest grid extended from $r_{\rm
  min}$ to $r=400M$, hence covering both the black hole and the initial scalar field distribution.
Following Ref.~\cite{Barranco:2012qs}, we study the spectrum by evaluating the Fourier transform in time of
$\phi(t,r)$ at many locations $r$, and then calculating the average. 

\begin{figure}[ht]
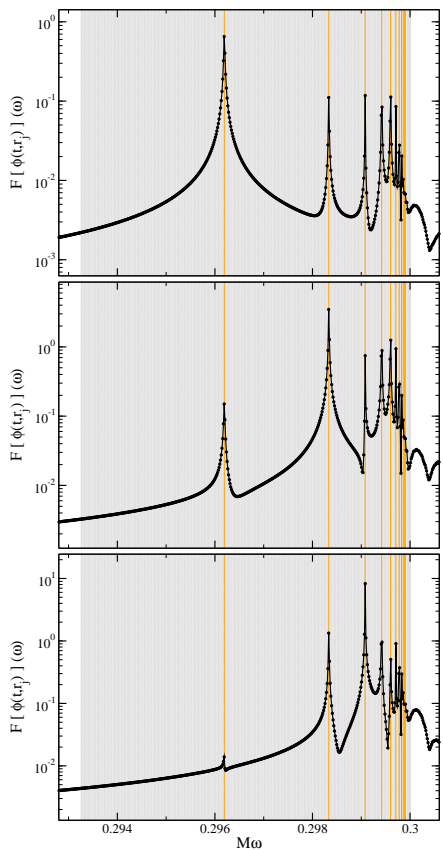

\begin{center}
\includegraphics[angle=0,width=0.32\textwidth,height=!,clip]{fourier_m0.3p_0-100.eps}
\includegraphics[angle=0,width=0.32\textwidth,height=!,clip]{fourier_m0.3p_100-200.eps}
\includegraphics[angle=0,width=0.32\textwidth,height=!,clip]{fourier_m0.3p_200-300.eps}
\end{center}
\caption{The Fourier spectrum for evolutions with initial scalar field configurations with
  $[R_1,R_2]=[0M,100M]$, $[100M,200M]$ and $[200M,300M]$. The gray shaded area
  indicates the resonance region, and the vertical lines indicate the modes'
  frequencies.
\label{f:fourier1} }
\end{figure}

\begin{figure}[ht]
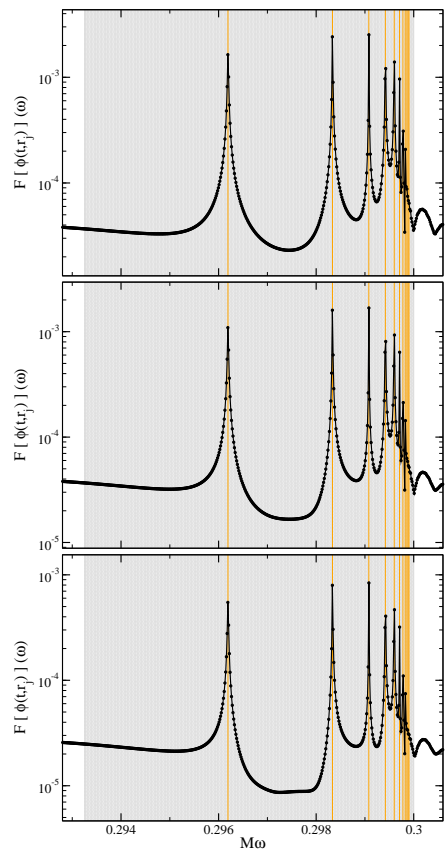

\begin{center}
\includegraphics[angle=0,width=0.32\textwidth,height=!,clip]{fourier_m0.3p_3-9.eps}
\includegraphics[angle=0,width=0.32\textwidth,height=!,clip]{fourier_m0.3p_4-8.eps}
\includegraphics[angle=0,width=0.32\textwidth,height=!,clip]{fourier_m0.3p_5-7.eps}
\end{center}
\caption{The Fourier spectrum for evolutions with initial scalar field configurations with
 $[R_1,R_2]=[3M,9M]$, $[4M,8M]$  and $[5M,7M]$. The gray shaded area
  indicates the resonance region, and the vertical lines indicate the modes'
  frequencies.
\label{f:fourier2} }
\end{figure}

The evolutions considered here correspond to initial configurations with
scalar field ``pulses''  with $[R_1,R_2]=[0,100M]$, $[100M,200M]$, $[200M,300M]$, $[3M,9M]$,
$[4M,8M]$ and $[5M,7M]$.
The spectra obtained from these evolutions
are shown in Figs.~\ref{f:fourier1} and~\ref{f:fourier2},
where we can see a perfect agreement between the peaks positions and the mode frequencies
(indicated by the vertical lines).  Note how the quasi-bound modes are exited even for
very ``thin'' initial scalar distributions, Fig.~\ref{f:fourier2},
although with smaller amplitudes.
Let us point out that the case with $[R_1,R_2]=[100M,200M]$, center panel of Fig.~\ref{f:fourier1},
  was already presented in~\cite{Barranco:2012qs}, with the difference that in
  that work there was some noise to the right of the resonant
  region (indicated in gray in the figure). In that occasion we claimed that the noise was originated at the
  right boundary, and that it did not notably affect the spectrum inside the
  resonant region. Here we were able to eliminate that noise using mesh
  refinement to push the right boundary far enough, showing that our previous
  claims were correct.

Given how generically the quasi-bound modes seem to appear in dynamical evolutions,
one may wonder how easy it really is to see them manifest as an effective scalar hair (or wig) in
general. Put another way, we can ask which conditions allow, and furthermore favor,
long lasting scalar field distributions around black holes. %
To answer this question, we first remind the reader that
quasi-bound modes on a Schwarzschild spacetime cannot exist unless the resonance condition 
\begin{eqnarray}
(M \mu)^2 &<& - \frac{1}{32}(\ell^2+\ell - 1)(\ell^2+\ell + 1)^2 \nonumber  \\
&+& \frac{1}{288}\sqrt{3(3\ell^4+6\ell^3+5\ell^2+2\ell+3)^3} \; 
\label{condition_real}
\end{eqnarray}
is satisfied~\cite{Burt:2011pv}. This represents a small range of $M\mu$
for typical values of $\ell$. For example, for $\ell=0$ the condition is $0<M\mu<1/4$. %in natural units. 
Second, even when the parameters are in the resonance region, depending on the
initial configuration, the quasi-bound modes may get excited with a very small
amplitude relative to the initial scalar field distribution. While for very wide initial configurations
most of the scalar field remains as a combination of quasi-bound modes, thinner
configurations (of order $M$ for instance) 
lose most of their mass by falling into the black hole, or being 
radiated towards infinity.
This difference between wide and thin initial configurations can be noted by comparing the amplitudes in
Fig.~\ref{f:fourier1} (wide initial configuration) with those in
Fig.~\ref{f:fourier2} (thin initial configuration).
Although the conditions leading to very long lasting scalar fields around black
holes are somehow restrictive, we would like to emphasize that they
may be satisfied in astrophysically relevant scenarios, as discussed
in Ref.~\cite{Barranco:2012qs}.

\subsection{Comparison with the semianalytical results}

Now we proceed to compare the numerical evolutions with the semianalytical results presented in
Sec.~\ref{Sec:KG}. For definiteness we consider two evolutions, corresponding
to $M\mu = 0.30$ and $M\mu = 0.20$, and 
in both cases we fix $\ell = 1$. The initial data is supported inside the interval $[R_1,R_2] = [4M,8M]$, see Eq.~(\ref{eq:id}) above. 
We compare the results of the evolution both in time and in spatial distribution. 

\begin{figure*}[ht]
\begin{center}
\includegraphics[angle=270,width=0.49\textwidth,height=!,clip]{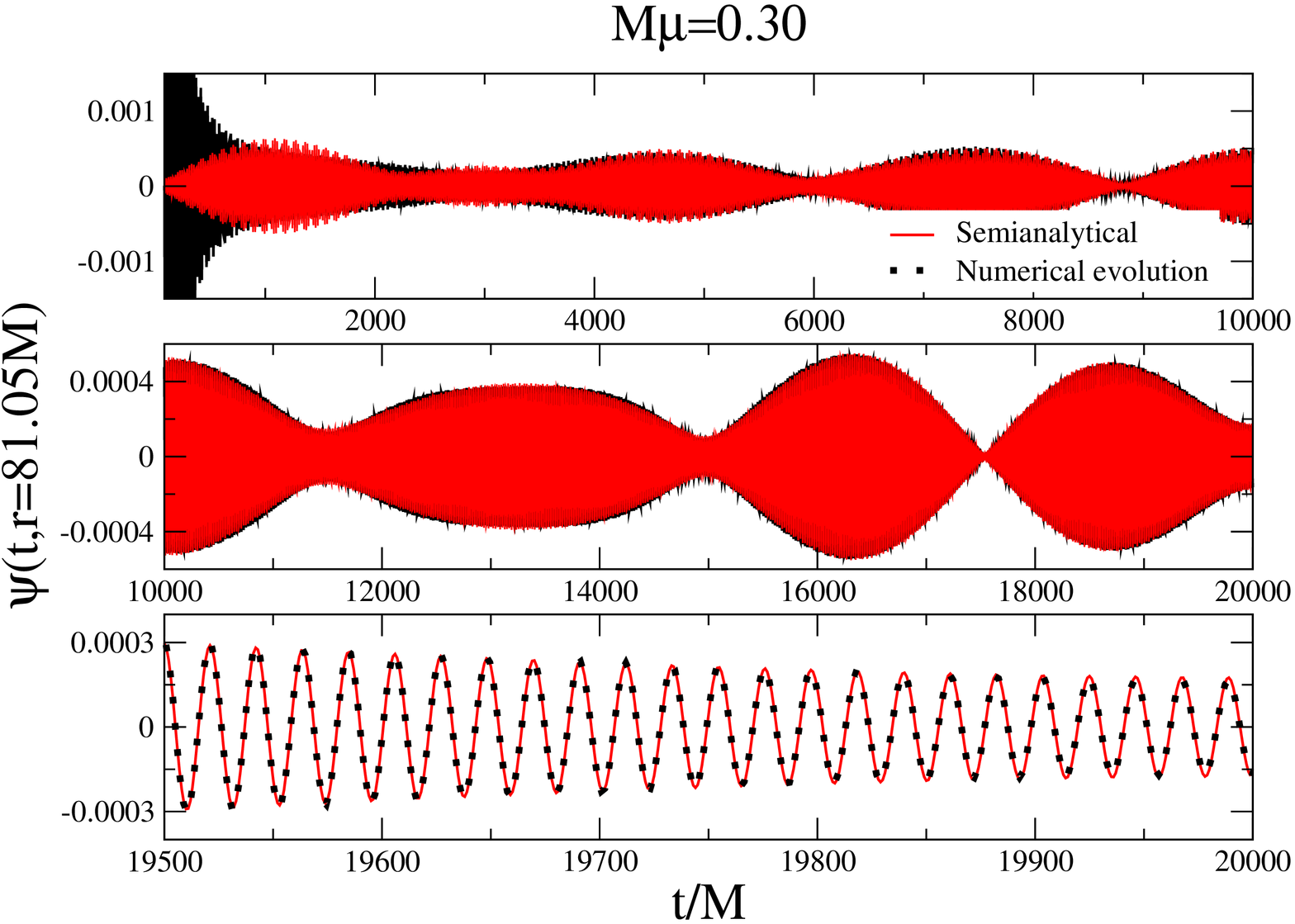}
\includegraphics[angle=270,width=0.49\textwidth,height=!,clip]{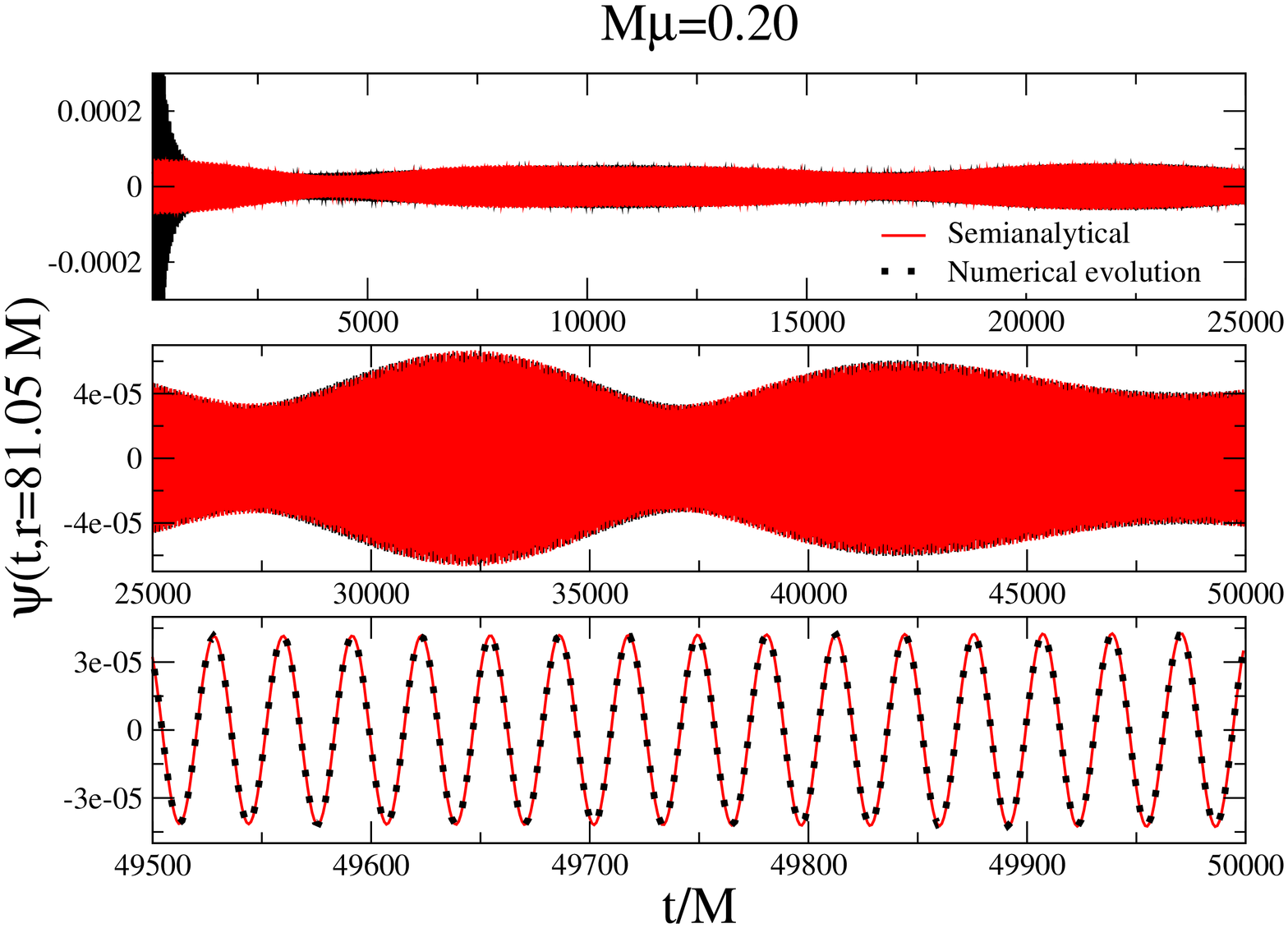}
\end{center}
\caption{The scalar field, $\psi(t,r)=\phi(t,r)/r$, as a function of time at $r=81.05M$ for initial data of the form in Eq.~(\ref{eq:id}), with $R_1=4M$ and $R_2=8M$.
Comparison between dynamical evolutions (dots) and semianalytical results
containing only the first nine modes (solid line). 
{\em Left panel:} $\ell=1$, $M\mu=0.3$.
{\em Right panel:} $\ell=1$, $M\mu=0.2$.
}
\label{Fig:Comp1} 
\end{figure*}

In order to compare as a function of time, we fix an observer at a point $r=81.05M$, although
we have verified that similar results appear for various other values of $r$.
These comparisons are shown in Fig.~\ref{Fig:Comp1}. 
As also pointed out in Ref.~\cite{Witek:2012tr}, we can see a ``beating'' due to the
combination of different quasi-bound frequencies.
We note that, for this particular case, evaluating the contribution from the
first nine quasi-bound modes was enough to have
a good matching with the dynamical evolution.

\begin{figure*}[ht]
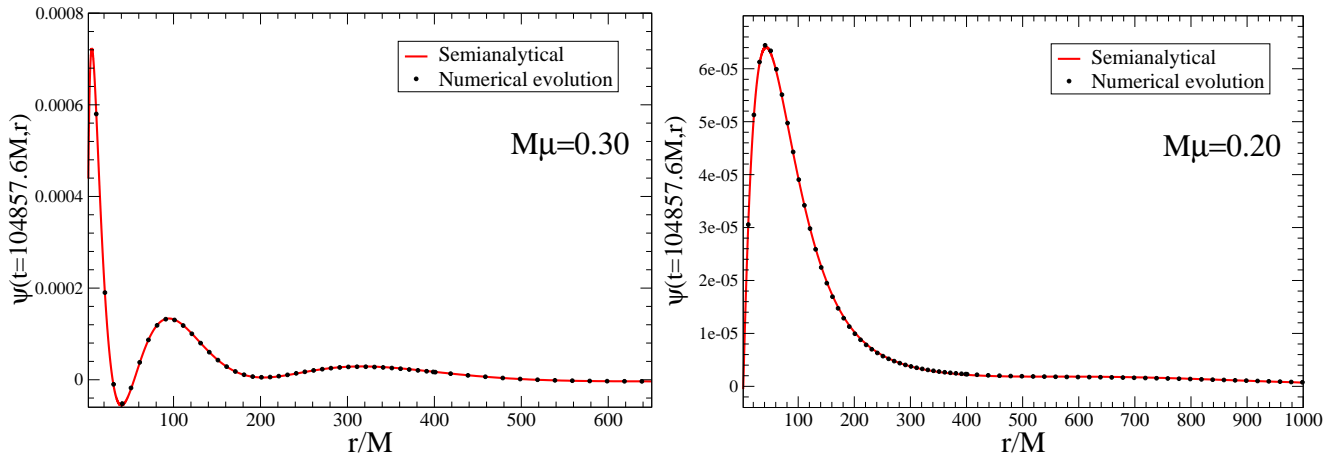

\begin{center}
\includegraphics[angle=0,width=0.48\textwidth,height=!,clip]{t3_mu30.eps}
\includegraphics[angle=0,width=0.49\textwidth,height=!,clip]{t3_mu20.eps}
\end{center}
\caption{
The scalar field, $\psi(t,r)=\phi(t,r)/r$, as a function of radius at $t=104857.6M$ for initial data of the form in Eq.~(\ref{eq:id}), with $R_1=4M$ and $R_2=8M$.
Comparison between numerical evolutions (dots) and semianalytical results
containing only the first nine modes (solid line). 
{\em Left panel:} $\ell=1$, $M\mu=0.3$.
{\em Right panel:} $\ell=1$, $M\mu=0.2$.
}
\label{Fig:Comp2} 
\end{figure*}

Concerning the distribution along the radial coordinate, we can
compare the semianalytical result at a given time with a snapshot of the numerical
evolution. In particular, we will take the function $\psi(t=104857.6M,r)$ obtained as the
superposition of the first nine quasi-bound modes shown in the lower panels of
Fig.~\ref{Fig:analytical}. The result of such comparison is presented in
Fig.~\ref{Fig:Comp2},
where we can see an excellent agreement between the two methods. 

\begin{figure}[ht]
\begin{center}
\includegraphics[angle=0,width=0.45\textwidth,height=!,clip]{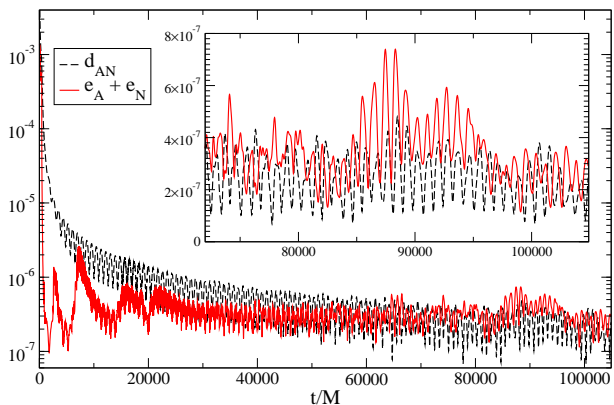}
\end{center}
\caption{The difference between semianalytical and
  numerical solutions, $d_{\rm AN}$, and their respective errors, $e_{\rm A}$
  and $e_{\rm N}$ are shown for the case $\ell=1$, $M\mu=0.2$.}
\label{Fig:errors} 
\end{figure}

To perform a more quantitative comparison it is convenient to first do an error
estimation for each case. For a semianalytical solution obtained from the
first $m$ modes,
$\Psi_{\rm A}(t,r)=\sum_{n=1}^m\Psi_n(t,r)$, we estimate the error as the last term in the
series. For a numerical evolution
 with resolution $\Delta r$, $\Psi_{\rm N}(t,r)=\Psi_{\Delta r}(t,r)$, we estimate the error
 as the difference with a solution with resolution 
$2\Delta r$. In order to do a comparison in time, we take the $L^2$
 norm in $r$ of the difference between the analytical
 and numerical solutions and their respective errors:
 $d_{\rm AN}(t):=||\Psi_{\rm A}(t,r)-\Psi_{\rm N}(t,r)||$, $e_{\rm
   A}(t):=||\Psi_m(t,r)||$, $e_{\rm N}(t):=||\Psi_{\Delta
   r}(t,r)-\Psi_{2\Delta r}(t,r)||$. We consider that a good agreement up to
 numerical error is achieved whenever $d_{\rm AN}\lesssim e_{\rm A}+e_{\rm
   N}$. These quantities are shown in Fig.~\ref{Fig:errors} for the same cases
 shown in the previous figures, that is $m=9$ and $\Delta r=0.1M$ in the inner
 grid. As expected, we see good agreement only at late times, when the
 semianalitycal solution becomes a good approximation. Note that these error
 estimations can be considered when deciding where to stop the semianalitical
 series. For the results presented in this work the error $e_{\rm A}$ is of the
 order of 1\% at the later times shown.

%%%%%%%%%%%%%%%%%%%%%%%%%%%%%%%%%%%%%%%%%%%%%
%%%%%%%%%%%%%%%%%%%%%%%%%%%%%%%%%%%%%%%%%%%%%
\section{Possible observational features}
\label{Sec:Discussion}
%%%%%%%%%%%%%%%%%%%%%%%%%%%%%%%%%%%%%%%%%%%%%
%%%%%%%%%%%%%%%%%%%%%%%%%%%%%%%%%%%%%%%%%%%%%

There has been some recent interest in the possibility that dark matter may be described by the coherent excitation of a massive scalar field~\cite{Turner:1983he,
Sin:1992bg, Peebles:1999fz, Peebles:2000yy, Sahni:1999qe, Hu:2000ke,
Matos:2000ng, Matos:2000ss, Arbey:2001qi, Matos:2003pe, Lee:2008jp,
Sikivie:2009qn, Marsh:2010wq, Lundgren:2010sp, Su:2010bj, Briscese:2011ka,
Harko:2011xw, Lora:2011yc, GonzalezMorales:2012uw, Rindler-Daller:2013zxa}.
In the axiverse~\cite{Arvanitaki:2009fg,Acharya:2010zx,Arvanitaki:2010sy,Marsh:2011gr,Kodama:2011zc,Macedo:2013qea,Marsh:2013taa,Tashiro:2013yea}, scalar fields with masses in 
the range $10^{-33}$ to $10^{-10}$ eV might naturally appear. 
It is very interesting that precision black hole physics can be used to explore such a possibility~\cite{Arvanitaki:2010sy,Kodama:2011zc,Macedo:2013qea}.
In this section we speculate about some further astrophysical observables that can be easily computed from the results in this paper.

The semianalytic method presented in Sec.~\ref{Sec:KG} (and tested against numerical evolutions in Sec.~\ref{Sec:Numerical}) offers interesting possibilities. 
Among them, it allows the evaluation of the scalar field at any time $t$ without the need of a full numerical evolution, as long as $t$ is large enough such that the contribution of
the high-frequency arc vanishes.\footnote{At very large times, the tail contribution to the integral kernel could in principle have some 
relevance~\cite{Koyama:2001ee,Koyama:2001qw}, but it does not appear for the time scales of our
simulations; see Sec.~\ref{subsec:numerical} for similar results
in the case of the toy model.} This allows us to study efficiently the amount of scalar field that ``survives'' around the black hole in the form of the 
superposition of quasi-bound modes that are excited by the initial data. We have referred to these remaining scalar field configurations as Schwarzschild scalar wigs~\cite{Barranco:2012qs}.
Furthermore, we can estimate the size of these wigs by estimating the size of each quasi-bound mode.
Note that even a narrow initial data, with a width of just $4M$, can excite a large number of modes. Most of them, as shown in Fig.~\ref{Fig:analytical}, have radial extension of hundreds or thousands
of $M$. 

\begin{figure}[ht]
\begin{center}
\includegraphics[angle=0,width=0.45\textwidth,height=!,clip]{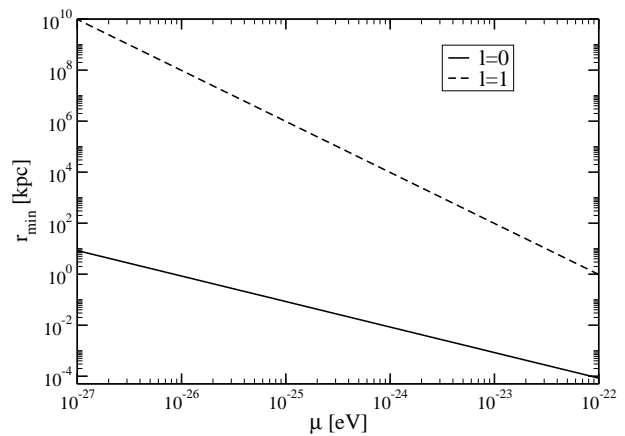}
\end{center}
\caption{Location, $r_{\rm min}$, of the local minimum of $V(r)$ for
  $M=10^8M_{\odot}$, $\ell = 0$ (continuous line) and $\ell = 1$ (dashed line), as a
  function of $\mu$. The value of $r_{\rm min}$ gives a 
  lower bound for the size of the scalar wigs.
\label{f:rmin} }
\end{figure}

In order to obtain a rough estimate on the size of the quasi-bound modes,\footnote{A more precise definition would be to determine the radius of a sphere
containing, let us say, 99\% of the mass.} we can calculate the location of the local minimum, $r_{\rm min}$, of the effective potential $V(r)$ defined in Eq.~(\ref{Eq:Veff}). Since the actual size is
usually much larger~\cite{Burt:2011pv}, this value gives a lower bound for the size of the scalar wig. Fig.~\ref{f:rmin} shows $r_{\rm min}$ as a function of $\mu$ for $\ell=0$ and
$\ell=1$, and black hole masses of order $M=10^8M_{\odot}$. 
Larger values of $\ell$ give larger values of $r_{\rm min}$. The
values shown in Fig.~\ref{f:rmin} seem to be consistent with galactic halos. For $\ell>0$ we obtain
sizes larger than those of the typical visible part of galaxies. For the case
with $\ell=0$, although the size seems too small, let us emphasize 
that $r_{\min}$ is usually much smaller than the actual size. Moreover, even if
the modes with $\ell=0$ were smaller than the typical size of actual galactic halos, in order to
describe a galaxy one would actually have to consider combinations of modes with
different values of $\ell$ for a more realistic scenario, hence yielding
larger sizes.

\begin{figure}[ht]
\begin{center}
\includegraphics[angle=0,width=0.48\textwidth,height=!,clip]{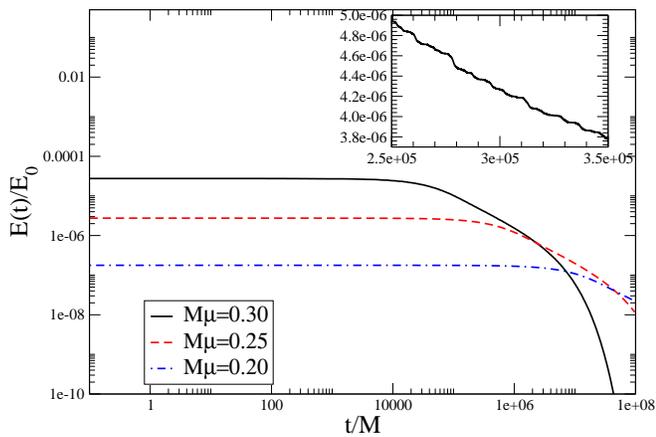}
\end{center}
\caption{Evolution of the ratio $E_{\textrm{modes}}(t)/E_0$ between the total energy contained in the mode part of the solution (first nine modes only) and the energy of the initial data as a function
of time. The initial
data is of the form given in Eq.~(\ref{eq:id}), with $R_1=4M$ and $R_2=8M$, and different values of $M\mu$. Note that $E_{\textrm{modes}}(0)/E_0\neq1$ because
Eq.~(\ref{Eq:energy2}) is not an appropriate representation of $E(t)$ at short times.
}
\label{Fig:E1} 
\end{figure}

Concerning the amount of scalar field in one of those wigs, we will compute now the total energy, $E(t)$, contained in a slice $t = \textrm{const}$~\cite{Burt:2011pv}. 
For the particular case of the Klein-Gordon equation in ingoing Eddington-Finkelstein coordinates it can be shown that
\begin{eqnarray}
E(t)&=&\int\limits_{2M}^\infty \left[ 
\left(1+\frac{2M}{r}\right)\left|\frac{\partial\phi}{\partial t}\right|^2
+ \left(1-\frac{2M}{r}\right)\left|\frac{\partial\phi}{\partial r}\right|^2 \right.
\nonumber \\
&& \qquad \left. 
 + \left(\frac{\ell(\ell+1)}{r^2}+\frac{2M}{r^3}+\mu^2\right)|\phi|^2 \right] dr\,.
\label{Eq:energy}
\end{eqnarray} 
At late times the scalar field can be written as a superposition of quasi-bound states, $\phi(t,r)=\sum_n A_n e^{s_n t}\phi_n(r)$, where the mode solutions 
in Eddington-Finkelstein coordinates $\phi_n(r)$ are related to the mode functions $f_{-}(s_n,x)$ defined in Eq.~(\ref{Eq:fmHeun}) through the relation $\phi_n(r) = (r/2M-1)^{-2Ms_n} f_-(s_n,z)$; see
Appendix~\ref{App:A}. Given the properties of the Heun
function, it follows that $\lim_{r\to 2M}\phi_n(r) = 1$. Using these identities and integration by parts, it can be shown that the energy in Eq.~(\ref{Eq:energy}) can be rewritten as
\begin{equation}
E_{\textrm{modes}}(t) =-2\sum_{n,m}e^{(s_n^*+s_m)t}\frac{s_n^*s_m}{s_n^*+s_m}A_n^*A_m\,.
\label{Eq:energy2}
\end{equation}
Therefore, at late times the total energy of the scalar field can be computed solely from the quasi-bound frequencies and the excitation amplitudes corresponding to a given initial data. For the
particular case of the initial data defined in Eq.~(\ref{eq:id}) (with $R_1 = 4M$ and $R_2 = 8M$), we have computed the energy $E_{\textrm{modes}}(t)$ given in 
Eq.~(\ref{Eq:energy2}) for large times, $t > 10,000M$, and compared it to the initial energy, $E_0$, which was computed from the integral in Eq.~(\ref{Eq:energy}). The evolution of the normalized
energy $E_{\textrm{modes}}(t)/E_0$ for the same initial data but different values of $M\mu$ is shown in Fig.~\ref{Fig:E1}.

The following comments on the behavior of $E_{\textrm{modes}}(t)/E_0$ are interesting. In contrast to the evolution of the energy shown in Ref.~\cite{Burt:2011pv}, where only one quasi-bound state was
evolved and an exponential decay was obtained, for the case of an arbitrary data the behavior of $E(t)/E_0$ shows transitions from one exponential decaying phase to another one. This is due to the
fact that
at late times, the solution is a superposition of different quasi-bound states. However, note that the energy $E_{\textrm{modes}}(t)$ is a non-increasing function of time, since the time derivative of
the right-hand side of Eq.~(\ref{Eq:energy2}) is equal to $-2$ times a perfect square. Another interesting property is that for given initial data, the
amount of energy of 
the scalar field is initially smaller but lasts for a larger time when $M\mu$ is decreased. The reason for this relies in the fact that when $M\mu \to 0$, the real parts of the complex frequencies
$s_n$, 
giving rise to the decay of
the solution, become very small, so that the configuration can last for a longer time. 

The profile in the energy density $\rho_{nlm}(t,r)$ of each individual mode $\phi_{nlm}$ is of the form shown in Figs.~6 and~7, Ref.~\cite{Burt:2011pv}.
However, the energy density of a combination of modes with different frequencies will have nontrivial
oscillations, due to its nonlinear dependency on the scalar field. Assuming
that galactic halos are described by a combination of 
these modes, such oscillations might be observable, 
producing changes in the galactic rotation curves over time. 

Let us consider a combination of modes with two frequencies, $\omega_1$ and
$\omega_2$. The energy density will be given by the sum of the energy density
of each mode (a quasi-static term), plus a cross term (given the quadratic form
of the density as function of the scalar field) with frequency
$\omega_r=\left|\omega_1-\omega_2\right|$. 

In general, the maximum possible frequency $\omega_r$ will
then be given by the difference of the higher and lower frequencies. This
value will be approximately equal to the upper bound given by the size of
the resonance region: 
${\omega_b}^2 = {\rm min} \left\{ \mu^2,V_{\rm max} \right\} - V_{\rm min}$,
where $V_{\rm max}$ ($V_{\rm min}$) is the local maximum (minimum) of the
effective potential $V(r)$.
In terms of periods, we have the lower bound $T_b=2\pi/\omega_b$. 

In order to give some estimations, if we choose  $M=10^{10}M_{\odot}$, $\mu=10^{-22}$~eV, and
$\ell=0$, we have $T_b=17$~years, thus in this case it seems plausible to detect the
oscillations by, for instance, comparing rotation curves observed during
a few decades of observation. This relatively small value of $T_b$ is obtained
for minimum $\ell$, and large $M$ and $\mu$, although still within acceptable
limits for galactic halos.
However, as soon as we choose slightly smaller $M$ or $\mu$, or
larger $\ell$, the period $T_b$ gets extremely larger. For example,
$M=10^8M_{\odot}$, $\mu=10^{-23}$~eV and $\ell=0$ gives
$T_b=5500$~years;  $M=10^5M_{\odot}$, $\mu=10^{-22}$~eV and $\ell=1$
gives $T_b=10^{11}$~years. Thus, it remains unclear whether the oscillations
might be observed.

%%%%%%%%%%%%%%%%%%%%%%%%%%%%%%%%%%%%%%%%%%%%
\section{Conclusions}
\label{sec:conclusions}
%%%%%%%%%%%%%%%%%%%%%%%%%%%%%%%%%%%%%%%%%%%%

We have considered the late time behavior of localized scalar field configurations surrounding a Schwarzschild black hole, using both numerical evolutions 
and the Green's function representation technique. At any time in the evolution, the scalar field can be represented as the convolution of the initial data with a suitable integral kernel. The
integral kernel can be divided into three different contributions, where each one is a contour integral  in the complex frequency plane over the Green's function of the problem. After some transient
initial period, and for the time scales reached with our numerical evolutions, we have shown that the scalar field can be accurately described by a superposition of quasi-bound modes alone, which
correspond to the residua of the poles of the Green's function.

Given arbitrary initial data for the scalar field on a spacelike hypersurface, the amplitude of each quasi-bound state in the solution can be obtained by computing a simple,
one-dimensional integral. Therefore, once these amplitudes have been calculated, it is possible to predict the late time behavior of the scalar field distribution without performing a numerical evolution.

These conclusions were reached by applying the Green's function technique first to a toy model consisting of a potential formed by a delta and a step function, where the modes can be computed in exact
form, and where the asymptotic form of the frequencies and the amplitudes can be obtained analytically. In particular, we showed that for this model problem all the poles of the Green's function are
located very close to the resonant band, and thus have very small decay rates. By comparison with a numerical evolution we showed that at late times the solution is accurately described by a
superposition of quasi-bound modes. Then, we turned to the case of a massive scalar 
field on a Schwarzschild black hole background, for which the quasi-bound frequencies, modes and amplitudes were determined semianalytically, and showed that in this case too the agreement with the
numerical evolutions at late times is remarkable.

We discussed some of the astrophysical implications of our findings, in particular to the scalar field dark matter models and to the axiverse. We corroborated the results in our
previous works, where we found that for appropriate values of the scalar field and the black hole mass the quasi-bound states can be extremely long lived. We also estimated the size of the wigs for
supermassive black holes, and speculated about a possible observational feature in the rotational curves of galaxies.

Although the method presented in this paper has been applied only to the case of a massive scalar field living on a Schwarzschild spacetime background, it can be naturally extended to 
higher order spin fields on other stationary spacetime geometries.

%%%%%%%%%%%%%%%%%%%%%%%%%%%
%%%   ACKNOWLEDGMENTS   %%%
%%%%%%%%%%%%%%%%%%%%%%%%%%%

\acknowledgments
The authors are grateful to Sam Dolan for helpful suggestions.
This work was supported in part by CONACyT through grants 82787, 101353,
167335, 182445, and by DGAPA-UNAM through grants IN115311 and IN103514.
AB and MM acknowledge support from CONACyT-M\'exico.
JCD also acknowledges support from CONACyT-M\'exico, and from FCT via project No. 
PTDC/FIS/116625/2010 and NRHEPÐ295189 FP7-PEOPLE-2011-IRSES.
OS was supported by a CIC grant to 
Universidad Michoacana de San Nicol\'as de Hidalgo. MM was also supported by a grant from the 
John Templeton Foundation. The opinions expressed in this publication are those of the authors 
and do not necessarily reflect the views of the John Templeton Foundation.

\appendix

%%%%%%%%%%%%%%%%%%%%%%%%%%%%%%%%%%%%%%%%%%%%
\section{Green's function representation for arbitrary time slices}
\label{App:A}
%%%%%%%%%%%%%%%%%%%%%%%%%%%%%%%%%%%%%%%%%%%%

In this appendix we generalize the integral representation~(\ref{Eq:IntegralRepresentation}) for the solutions of Eq.~(\ref{Eq:Cauchy1}) to arbitrary spacelike foliations $\overline{t} =
\textrm{const.}$,
where the new time coordinate $\overline{t}$ parametrizing the new foliation is related to $t$ through
\begin{equation}
\overline{t} = t - h(x),
\label{Eq:tbar}
\end{equation}
with $h: \Real \to \Real$ the height function. $h$ should be smooth and satisfy $|h'(x)| < 1$ in order to ensure that the time slices $\overline{t} = \textrm{const.}$ are spacelike. With respect to
the new
coordinates $(\overline{t},r)$ Eq.~(\ref{Eq:Cauchy1}) is transformed into the following equation for $\overline{\phi}(\overline{t},x) = \phi(t,x)$.
\begin{equation*}
\frac{\partial^2\overline{\phi}}{\partial\overline{t}^2} 
 - \left( \frac{\partial}{\partial x} - h'(x)\frac{\partial}{\partial\overline{t} } \right)^2
 \overline{\phi} + V(x)\overline{\phi} = 0.
\end{equation*}
We solve this equation given initial data on the initial surface $\overline{t}=0$ which is of the form
\begin{equation*}
\overline{\phi}(0,x) = \overline{\phi_0}(x),\quad 
\frac{\partial\overline{\phi}}{\partial t}(0,x) = \overline{\pi_0}(x),\quad x\in \Real.
\end{equation*}
To do this, we perform a Laplace transformation in time, which yields the ordinary differential equation
\begin{equation}
\left[ s^2 - \left( \frac{\partial}{\partial x} - h'(x) s \right)^2 + V(x) \right]\tilde{\phi}(s,x)
 = F(s,x),
\label{Eq:Laplace}
\end{equation}
where $\tilde{\phi}(s,x)$ refers to the Laplace transformation of $\overline{\phi}(\overline{t},x)$ and the source term is given by
\begin{eqnarray*}
F(s,x) &=& (1-h'(x)^2)(\overline{\pi_0}(x) + s\overline{\phi_0}(x) ) \nonumber\\
 &+& 2h'(x)\frac{\partial\overline{\phi_0}}{\partial x}(x) + h''(x)\overline{\phi_0}(x).
\end{eqnarray*}
The solution of Eq.~(\ref{Eq:Laplace}) which is well-behaved at $x\to \pm\infty$ can be written as
\begin{equation*}
\tilde{\phi}(s,x) = \int\limits_{-\infty}^\infty \overline{G}(s,x,y) F(s,y) dy,
\end{equation*}
with the Green's function
\begin{equation}
\overline{G}(s,x,y) := \frac{1}{\overline{W}(s,y)}\left\{ \begin{array}{ll} 
 \overline{f_-}(s,y)\overline{f_+}(s,x),  & y\leq x \\ 
 \overline{f_-}(s,x)\overline{f_+}(s,y), & y > x 
\end{array} \right\},
\label{Eq:GreenFctBis}
\end{equation}
where here $\overline{f_-}(s,x)$ and $\overline{f_+}(s,x)$ are two nontrivial homogeneous solutions of Eq.~(\ref{Eq:Laplace}), with $\overline{f_-}$ 
representing an outgoing mode at $x\to -\infty$ and
$\overline{f_+}$ and outgoing mode at $x\to +\infty$. $\overline{W}(s,x)$ is the Wronski determinant of $\overline{f}_\pm$, defined as
\begin{equation*}
\overline{W}(s,x) := \det\left( \begin{array}{ll}
 \overline{f_+}(s,x) & \overline{f_-}(s,x) \\
 \overline{f_+}'(s,x) & \overline{f_-}'(s,x)
\end{array} \right).
\end{equation*}

What is the relation between $\overline{G}(s,x,y)$ and the Green's function $G(t,x,y)$ corresponding to the original time slices $t= \textrm{const.}$? Because of Eq.~(\ref{Eq:tbar}) the mode solutions
$\overline{f_\pm}$ and $f_\pm$ are related to each other through
\begin{equation*}
\overline{f_\pm}(s,x) = e^{s h(x)} f_\pm(s,x).
\end{equation*}
Accordingly, it follows that $\overline{W}(s,x) = e^{2sh(x)} W(s)$ and $\overline{G}(s,x,y) = e^{s(h(x)-h(y))} G(s,x,y)$. With these observations, it is not difficult to generalize the
formulae~(\ref{Eq:mode.amplitude}) to the case where the initial data is specified on the $\bar{t} = 0$ surface instead of the $t=0$ surface. The result is
\begin{equation}
\overline{\phi}_{\textrm{modes},n}(\bar{t},x) = \overline{A_n} e^{s_n t} f_+(s_n,x)
 = \overline{A_n} e^{s_n\overline{t}}\overline{f_+}(s_n,x),
\label{Eq:ModeSolutionBis}
\end{equation}
with the amplitude
\begin{equation}
\overline{A_n} := \left. \left[ \frac{d}{ds} W(s) \right]^{-1} \right|_{s=s_n}
\int\limits_{-\infty}^\infty f_-(s_n,y) e^{-s_n h(y)} F(s_n,y) dy.
\label{Eq:AmplitudeBis}
\end{equation}
Comparing with Eqs.~(\ref{Eq:mode.amplitude}) we see that the only change required for computing the mode solution is the replacement $[s_n\phi_0(y) + \pi_0(y)] \mapsto e^{-s_n h(y)}
F(s_n,y)$ in the expression for the amplitude $A_n$.

As an example, for ingoing Eddington-Finkelstein coordinates we have with $r > 2M$,
\begin{equation*}
x = r + 2M\log\left( \frac{r}{2M} - 1 \right),\qquad
h(x) = r - x,
\end{equation*}
and
\begin{eqnarray*}
&& F(s,x) dx \\
&=& \left[ \left( 1 + \frac{2M}{r} \right)(\overline{\pi_0} + s\overline{\phi_0})
 - \frac{4M}{r}\frac{\partial\overline{\phi_0}}{\partial r} + \frac{2M}{r^2}\overline{\phi_0}
 \right] dr. \nonumber
\end{eqnarray*}
%

%%%%%%%%%%%%%%%%%%%%%%%%%%%%%%%%%%%%%%%%%%%%
\section{Contributions from the high-frequency arc and the branch cut to the solution of the toy model}
\label{App:B}
%%%%%%%%%%%%%%%%%%%%%%%%%%%%%%%%%%%%%%%%%%%%

In Sec.~\ref{Sec:Toy} we provided an analytic derivation of the mode part of the solutions to the toy model problem. In this appendix, we briefly discuss some properties of the remaining parts of
the solution, namely the ones coming from the integration over the high-frequency arc $\mu_R$ and the two curves $\Gamma_R$ around the branch cut (see Fig.~\ref{f:contour}). The corresponding integral
kernels are given by
\begin{equation*}
k_{\textrm{tail}}(t,x,y) = \lim\limits_{R\to\infty}
\frac{1}{2\pi i}\int\limits_{\Gamma_R} e^{st} G(s,x,y) ds,
\end{equation*}
and
\begin{equation*}
k_{\textrm{hfa}}(t,x,y) = \lim\limits_{R\to\infty}
\frac{1}{2\pi i}\int\limits_{\mu_R} e^{st} G(s,x,y) ds.
\end{equation*}
In terms of the complex angle $\varphi$ defined in Eq.~(\ref{Eq:varphi}) we can parametrized the integration curves $\Gamma_R$ and $\mu_R$ as follows. Let $\varepsilon > 0$ and $R > 0$. Then,
$\Gamma_R$ is defined as the union of the two hyperbolae $\varphi = \alpha + i\beta$ with $-R\leq \beta\leq R$ and $\alpha = \pm(\pi/2 - \varepsilon)$, respectively, and $\mu_R$ as the elliptic arc
described $\varphi = \alpha + i R$ with $-\pi/2 + \varepsilon \leq \alpha \leq \pi/2 - \varepsilon$, see Fig.~\ref{f:newcoords}. As in Sec.~\ref{Sec:Toy}, we assume that the initial data is
supported inside the potential well, $0 < x < a$, and we only consider the solution inside this region. Then, the Green's function $G(s,x,y)$ can be computed using
Eqs.~(\ref{Eq:GreenFct},\ref{Eq:c-}--\ref{Eq:d+}). In terms of the complex angle $\varphi$ one obtains:
\begin{widetext}
\begin{eqnarray*}
e^{st} G(s,x,y) ds = \frac{
 (1 + \lambda i\sin\varphi) e^{i\sin\varphi(\tau-\eta)} 
+ e^{i\sin\varphi(\tau + \eta - 2q) - 2i\varphi} 
- (1 + \lambda i\sin\varphi) e^{i\sin\varphi(\tau + \xi - 2q) - 2i\varphi}
- e^{i\sin\varphi(\tau-\xi)} }
 {2 \tan(\varphi) F(\lambda,\varphi)} d\varphi,
\end{eqnarray*}
\end{widetext}
where we have introduced the dimensionless quantities $\tau := \mu t$, $\xi := \mu(x+y)$, $\eta := \mu|x-y|$, and where we recall the definitions $\lambda := 2\mu/A$ and $q:=a\mu$ and the function
$F(\lambda,\varphi)$ defined in Eq.~(\ref{Eq:FDef2}).

Explicit evaluation of the resulting integrals is, of course, a non-trivial task. However, we prove the following result which shows that the contribution from the high-frequency arc vanishes
identically after two crossing times of the potential well:
\begin{proposition}
Let $t > 2a$. Then, $k_{\textrm{hfa}}(t,x,y) = 0$ for all $0 < x,y < a$.
\end{proposition}

\proof
We start with a basic estimate for the function 
$F(\lambda,\varphi) = 1 + i\lambda\sin(\varphi) - e^{-2i\varphi - 2qi\sin\varphi}$. Writing $\varphi = \alpha + i\beta$ 
with $-\pi/2 < \alpha < \pi/2$ and $\beta > 0$ we have
\begin{displaymath}
i\sin\varphi = -\cos\alpha\sinh\beta + i\sin\alpha\cosh\beta, \nonumber
\end{displaymath}
which has negative real part. Using the estimates $\sinh\beta \leq e^\beta$ and $\cosh\beta \leq e^\beta$ 
which are valid for $\beta > 0$ we obtain from this
\begin{displaymath}
|i\sin\varphi| = \sqrt{\cos^2\alpha\sinh^2\beta + \sin^2\alpha\cosh^2\beta} 
 \leq e^{\beta}. \nonumber
\end{displaymath}
On the other hand,
\begin{displaymath}
-2i\varphi - 2qi\sin\varphi = 2\beta + 2q\cos\alpha\sinh\beta 
- 2i(\alpha + q\sin\alpha\cosh\beta) \nonumber
\end{displaymath}
which implies
\begin{displaymath}
|e^{-2i\varphi - 2qi\sin\varphi}| = e^{2\beta + 2q\cos\alpha\sinh\beta} \geq e^{2\beta}. \nonumber
\end{displaymath}
Therefore, if follows that
\begin{equation}
|F(\lambda,\varphi)| \geq e^{2\beta} - |1 + i\lambda\sin\varphi|
 \geq e^{2\beta} - \lambda e^\beta - 1,
\label{Eq:FEstimate}
\end{equation}
and choosing $R > 0$ large enough, we have $|F(\lambda,\varphi)| \geq e^{2\beta}/2$ for all $\beta \geq R$. In particular, this proves that there are no poles which lie arbitrarily far from the origin
in the complex plane.

Next, we observe that
\begin{displaymath}
\frac{1}{|\tan^2(\varphi)|} 
 = \frac{\cosh^2\beta - \sin^2\alpha}{\sinh^2\beta + \sin^2\alpha}
 \leq \frac{\cosh^2\beta}{\sinh^2\beta}. \nonumber
\end{displaymath}
The right-hand side converges to one for $\beta\to\infty$, implying that $1/|\tan(\varphi)|$ is uniformly bounded for large $\beta$'s.

Now that we have control on the terms in the denominator, it is not difficult to estimate each term of the Green's function. For example, for the first term we find
\begin{eqnarray*}
&& \left| \int\limits_{\mu_R}
\frac{ (1 + \lambda i\sin\varphi) e^{i\sin\varphi(\tau-\eta)} }
 {2 \tan(\varphi) F(\lambda,\varphi)} d\varphi \right| \\
 &\leq& \textrm{const}\times
 e^R e^{-2R} \int\limits_{-\pi/2+\varepsilon}^{\pi/2-\varepsilon}
 e^{-\cos\alpha \sinh(R)(\tau-\eta)} d\alpha\\
 &\underset{R\to\infty}{\longrightarrow}& 0,
\end{eqnarray*}
since $\tau-\eta = \mu(t - |x-y|) > 0$. The second and fourth terms can be estimated in a similar way. For the third term we obtain
\begin{eqnarray*}
&& \left| \int\limits_{\mu_R}
\frac{ (1 + \lambda i\sin\varphi) e^{i\sin\varphi(\tau+\xi-2q) - 2i\varphi} }
 {2 \tan(\varphi) F(\lambda,\varphi)} d\varphi \right| \\
 &\leq& \textrm{const}\times
 e^R e^{-2R} \int\limits_{-\pi/2+\varepsilon}^{\pi/2-\varepsilon}
 e^{-\cos\alpha \sinh(R)(\tau+\xi-2q) + 2R} d\alpha\\
 &\leq& \textrm{const}\times 2\sinh(R)\int\limits_{-\pi/2+\varepsilon}^{\pi/2-\varepsilon}
 e^{-\cos\alpha \sinh(R)(\tau+\xi-2q)} d\alpha.
\end{eqnarray*}
In the limit $R\to\infty$ this term also converges to zero. This can be seen by setting $z:=\sinh(R)(\tau+\xi - 2q) = \sinh(R)\mu(t+x+y-2a) > 0$ and using the estimate $\cos\alpha \geq 1 -
2\alpha/\pi$, $\alpha\geq 0$, such that
\begin{eqnarray*}
z \int\limits_{-\pi/2+\varepsilon}^{\pi/2-\varepsilon} e^{-z\cos\alpha} d\alpha
 &\leq& 2z\int\limits_0^{\pi/2-\varepsilon} e^{-z(1 - 2\alpha/\pi)} d\alpha\\
 &=& \pi\left( e^{-2\varepsilon z/\pi} - e^{-z} \right) 
 \underset{R\to\infty}{\longrightarrow} 0.
\end{eqnarray*}
This concludes the proof of the proposition.
\qed

%%%%%%%%%%%%%%%%
%%%   REFERENCES   %%%
%%%%%%%%%%%%%%%%

%\bibliography{bibtex/referencias}
%\bibliographystyle{unsrt} 

\bibliographystyle{prsty}
\bibliography{referencias}
\bibliographystyle{unsrt}

\end{document}